\documentclass[a4paper,11pt]{article}
\pdfoutput=1 

\usepackage{jheppub} 
\usepackage{siunitx}

\usepackage[T1]{fontenc} 
\usepackage{csquotes}
\usepackage{import}
\usepackage{physics}
\usepackage{siunitx}
\usepackage{graphics}
\usepackage{caption}
\usepackage{subcaption}
\usepackage{collcell}
\usepackage{colortbl}
\usepackage{listings}
\usepackage{float}
\usepackage{cleveref}
\Crefformat{figure}{Fig.\,#2#1#3}
\Crefmultiformat{figure}{Figs.\,#2#1#3}{ and~#2#1#3}{, #2#1#3}{ and~#2#1#3}
\crefrangeformat{figure}{Figs.\,#3#1#4 to~#5#2#6}
\Crefformat{section}{Sec.\,#2#1#3}
\Crefformat{equation}{Eq.\,(#2#1#3)}
\usepackage[acronym]{glossaries}

\usepackage{placeins}

\newacronym{FKS}{FKS}{Frixione-Kunst-Signer}
\newacronym{DGLAP}{DGLAP}{Dokshitzer–Gribov–Lipatov–Altarelli–Parisi}

\newacronym{LHC}{LHC}{Large Hadron Collider}
\newacronym{RHIC}{RHIC}{Relativistic Heavy Ion Collider}

\newacronym{PDF}{PDF}{parton distribution function}
\newacronym{FF}{FF}{fragmentation function}
\newacronym{PS}{PS}{parton shower}
\newacronym{POWHEG}{POWHEG}{Positive Weight Hardest Emission Generator}

\newacronym{IR}{IR}{infrared}
\newacronym{UV}{UV}{ultraviolet}
\newacronym{IRC}{IRC}{infrared collinear}

\newacronym{LO}{LO}{leading-order}
\newacronym{NLO}{NLO}{next-to-leading-order}
\newacronym{NLL}{NLL}{next-to-leading-logarithmic}
\newacronym{NNLO}{NNLO}{next-to-next-to-leading-order}
\newacronym{MiNLO}{MiNLO}{Multi-Scale Next-to-Leading Order}

\newacronym{QGP}{QGP}{quark-gluon plasma}
\newacronym{QCD}{QCD}{quantum chromodynamics}
\newacronym{QED}{QED}{quantum electrodynamics}

\title{Prompt photon production with two jets in POWHEG}

\author[a]{Tom\'a\v{s} Je\v{z}o}
\author[a]{\!\!, Michael Klasen}
\author[a]{and Alexander Neuwirth}

\affiliation[a]{Institut  für  Theoretische  Physik,  Universität  Münster,  Wilhelm-Klemm-Straße 9, 48149 Münster, Germany}

\emailAdd{tomas.jezo@uni-muenster.de}
\emailAdd{michael.klasen@uni-muenster.de}
\emailAdd{alexander.neuwirth@uni-muenster.de}

\abstract{
    Prompt photon production is highly sensitive to the distribution of quarks and gluons in free protons and nuclei and an important baseline for phenomenological studies of the properties of the quark-gluon plasma. In this paper, we present a new calculation of the production of prompt photons in association with two jets at next-to-leading order in quantum chromodynamics matched to parton showers with the POWHEG method. This calculation extends our previous analysis of prompt photon production in association with one jet using \texttt{POWHEG+PYTHIA}. We investigate the role of the parton shower as an alternative description of the parton-to-photon fragmentation process and analyse correlations between the photon and the jets. In addition, we compare \texttt{POWHEG+PYTHIA} with \texttt{POWHEG+HERWIG} predictions, experimental ATLAS data for isolated photons, and a perturbative calculation at next-to-next-to-leading order. Both parton shower models bring the next-to-leading order prediction into good agreement with the experimental data and the next-to-next-to-leading order calculation.
}

\preprint{MS-TP-24-19}


\begin{document}
\maketitle
\flushbottom

\section{Introduction}
\label{sec:intro}

Photons are invaluable probes in the study of heavy-ion collisions thanks to their ability to traverse the dense medium without undergoing any final-state interactions. Generated at all stages of the collision, they originate from various processes including hard-scattering interactions such as quark-gluon Compton scattering and quark-antiquark annihilation, decays of light and heavy flavour hadrons, and thermal radiation from the hot medium produced during the collision. This makes photons sensitive indicators of the presence of a \gls*{QGP} and its properties \cite{Rapp:2009yu,Tserruya:2009zt,Wilde:2012wc,Reygers:2022crp,Andronic:2024rfn}.

Processes involving prompt photons, such as their production in association with jets, also provide a unique opportunity to study the distribution of gluons in free protons and nuclei due to their production mechanism via quark-gluon Compton scattering ($gq \to \gamma q$) \cite{Klasen:2013mga,Klasen:2014xra}. The gluon density becomes particularly large at low values of Bjorken-$x\sim 2 p_T^\gamma/\sqrt{s}$, where at some point saturation may occur \cite{Jalilian-Marian:1996mkd,Armesto:2022mxy}. The knowledge of precise distributions of parton momenta within a nucleon or nucleus, described in the framework of collinear factorisation by \glspl*{PDF}, is crucial for precise simulations of high-energy collisions.

Prompt photons can be produced through direct hard processes or fragmentation processes. Direct photons are generated in the hard interaction. Fragmentation photons, instead, are produced when a high-energy parton undergoes radiation, leading to a photon that is accompanied by other hadronic activity \cite{Koller:1978kq,Laermann:1982jr}. The distinction between these two types is critical and is often facilitated through isolation criteria in experimental setups.

Isolation techniques are also implemented to separate prompt photons from those produced in hadronic decays. They involve restricting the hadronic activity around a photon candidate, ensuring the hadronic energy in a cone around the photon is below a certain threshold \cite{Frixione:1998jh,Catani:2002ny,Amoroso:2020lgh}. These criteria help in minimising the contamination from fragmentation photons, allowing for a clearer interpretation of direct photon production. The effect of the isolation cone on the parton-to-photon fragmentation contribution has been thoroughly examined, for example, using \texttt{JETPHOX}~\cite{Belghobsi:2009hx,dEnterria:2012kvo}. Especially, in the low transverse momentum regime, the isolation cone is crucial to suppress the dominant fragmentation contribution compared to the photons originating in the hard process. To consistently eliminate the fragmentation contributions, one must exclude collinear parton-photon configurations without disrupting the cancellation of infrared singularities from soft gluon emissions \cite{Frixione:1998hn}. While straightforwardly vetoing the collinear configurations, as in the fixed cone approach, is not \gls*{IRC} safe, suppressing them maintains the necessary cancellation. This method is known as smooth cone isolation. The proof of infrared safety is similar to that of jet definitions, where analogous cone algorithms are also sensitive to \gls*{IRC} safety \cite{Atkin:2015msa}.

Early prompt photon calculations were performed at the next-to-leading logarithmic accuracy without the inclusion of fragmentation \cite{Baer:1990ra} and later at \gls*{NLO} in the small-cone approximation \cite{Gordon:1994ut} relying on Monte Carlo methods.
Following them, a new calculation released under the name of \texttt{JETPHOX} \cite{Catani:2002ny,Aurenche:2006vj,Belghobsi:2009hx} was published improving on the aforementioned shortcomings by including the direct and fragmentation contributions without any approximations.
\Gls*{NNLO} calculations for direct photon production have been performed in \texttt{MCFM} \cite{Campbell:2016lzl,Campbell:2017dqk,Campbell:2018wfu} and \texttt{NNLOJET} \cite{Chen:2019zmr,Schuermann:2022qdm}, but are unfortunately not yet publicly available.
Complementary to the above calculations, the \gls*{NLO}  \acrshort*{POWHEG} (\acrlong*{POWHEG})\glsunset{POWHEG} implementation in Ref.~\cite{Jezo:2016ypn} does not rely on fragmentation functions, but instead consistently attaches a \gls*{PS} to the generated events, allowing for more realistic studies of photon-jet correlations and photon fragmentation. Additionally, merging techniques, such as those used in the \texttt{Sherpa} framework, have advanced the accuracy of isolated photon simulations by incorporating multiple parton emissions described by exact matrix elements \cite{Gleisberg:2008ta,Sherpa:2019gpd}. \texttt{Sherpa} achieves \gls*{NLO} \gls*{QCD} matched to \gls*{PS} precision for the $2\to 3$ production process of a photon with two jets. 
Recently, even higher-order corrections to $\gamma jj $ production at NNLO has been achieved, further enhancing the precision of theoretical predictions \cite{Badger:2023mgf}.

We present a new calculation of prompt photon production with two jets at \gls*{NLO} \gls*{QCD} matched to \glspl*{PS} using the \gls*{POWHEG} method \cite{Nason:2004rx,Frixione:2007vw}. Extending the existing \texttt{POWHEG\,BOX}~\cite{Alioli:2010xd} direct photon calculation \cite{Jezo:2016ypn} by one jet is a significant advancement for several reasons. Firstly, incorporating the second jet at \gls*{NLO} improves the accuracy of the photon plus jet calculations by providing a more precise depiction of the event, reducing theoretical uncertainties. It also enables the study of jet-jet correlations within this process, offering new insights into the dynamics between jets in the presence of a direct photon. Moreover, the simulation allows for a detailed analysis of scenarios where a photon traverses a jet, including cases where the photon lies within the jet cone, which is crucial for understanding jet-photon interactions and isolation criteria. Lastly, this calculation enhances the capability to examine the ratio of $Z$+jets to $\gamma$+jets, thus improving background estimations in searches for new physics in $Z \to \nu \bar\nu$ decays \cite{CMS:2021fxy}.

There have been many experimental studies of isolated prompt photon production at the SppS, Tevatron, \gls*{RHIC} and the \gls*{LHC} as outlined in Ref.~\cite{Jonas:2023akr}. PHENIX \cite{PHENIX:2006duh} and ALICE \cite{ALICE:2019rtd} not only measured low transverse momenta ($\lesssim \SI{100}{GeV}$) direct photons, but they also studied low invariant mass dilepton pairs originating from virtual photons \cite{PHENIX:2009gyd,ALICE:2018fvj}. Typically, the yield of dielectrons with $m_{ee}\leq \SI{1}{GeV}$ has been computed by applying the Kroll-Wada prescription to real photon production \cite{Kroll:1955zu}. ATLAS \cite{ATLAS:2019buk} and CMS \cite{CMS:2018qao} extended these measurements up to the TeV scale and to PbPb collisions \cite{ATLAS:2023wzj}. In our phenomenological analysis, we will focus on a comparison of our predictions with the ATLAS measurement of isolated-photon plus two-jet production in $pp$ collisions at $\sqrt s=13$ TeV~\cite{ATLAS:2019iaa}.

The manuscript is structured as follows. We first discuss the \gls*{POWHEG} method and its application to prompt photon production in association with two jets at \gls*{NLO} in \Cref{sec:prompt}. Then we present our simulation setup and discuss the results of our calculations in \Cref{sec:res}. Finally, we summarise our findings and conclude in \Cref{sec:conc}.

\section{Prompt photon production in association with two jets at NLO}
\label{sec:prompt}

Building upon our previous work on prompt photon production in association with a single jet at NLO \cite{Jezo:2016ypn} in the {\tt POWHEG\,BOX\,V2} framework, we now perform the calculation of the process with an additional jet. The concepts and methods used in the calculation are similar to the ones used in the single jet case. However, the complexity of the calculation increases with the number of final state particles. In the subsequent sections, we will review the POWHEG method, detail the calculation of the process $pp \to \gamma jj$ at \gls*{NLO}, and describe the implementation of this calculation in {\tt POWHEG\,BOX\,V2}. Particular attention will be given to differences with respect to our previous calculation.

\subsection{The POWHEG method as an alternative description of fragmentation}
\label{sec:powheg}
In the context of prompt photon production, photon \glspl*{FF} play a crucial role. Analogous to \glspl*{PDF} $f_{a/A}$ in the initial state, the \glspl*{FF} $D_{\gamma/c}$ account for the non-perturbative long distance contributions in the final state. Similar to \glspl*{PDF} the \glspl*{FF} first need to be determined from experimental data such as prompt photon production in $e^+e^-$ collisions \cite{Klasen:2002xb,Nason:1997nu}, in hadronic collisions \cite{PHENIX:2012krx} as well as in vector meson production, where the photon is assumed to fluctuate into hadronic states \cite{Klasen:2014xfa}. Just like \glspl*{PDF}, \glspl*{FF} can be evolved in scale using the \gls*{DGLAP} evolution equations. However, in the case of photon \glspl*{FF}, the gluon characteristics in the evolution equations must be supplemented with those of the photons. Specifically, this involves replacing the gluon-related Casimir operators in the splitting kernels with those appropriate for photons and the quark-related Casimir operator by the electromagnetic charge of the quark. The factorised photon production cross-section with an untagged hadronic final state $X$ then reads
\begin{equation}
    \begin{split}
    \dd\sigma_{AB \to \gamma X} (p_A,p_B,p_\gamma)
    =&
    \sum_{a,b,c}\int_0^1\dd{x_a} \dd{x_b} \dd{z}
    f_{a/A}(x_a,\mu_{F_i})f_{b/B}(x_b,\mu_{F_i})
    \\
    &\dd \hat \sigma_{ab \to cX}(x_aP_A,x_bP_B,\frac{P_\gamma}{z},\mu_{R},\mu_{F_i},\mu_{F_f})
    D_{\gamma/c}(z,\mu_{F_f})
    \,,
    \end{split}
\end{equation}
where one sums over the intermediate initial and final state partons and convolves the \glspl*{PDF} $f_{a/A}$, picking parton $a$ from hadron $A$, and \glspl*{FF} $D_{\gamma/c}$, fragmenting parton $c$ to a photon, with the partonic cross-section $\dd \hat \sigma_{ab \to cX}$. These elements are connected through the renormalisation scale $\mu_R$ and factorisation scale in the initial state $\mu_{F_i}$ and final state $\mu_{F_f}$. The partonic cross-section $\hat \sigma$ is not \gls*{IRC} safe for identified partons. However, factorisation theorems ensure the cancellation of the remaining universal collinear divergences~\cite{Collins:1989gx}. Alternatively one can use a parton shower to model the fragmentation, which is the path we pursue in this study.
Advantages of the parton shower include more flexibility, customised tuning for \gls*{QGP} effects like jet quenching and energy loss and simulations beyond the inclusive transition to a final state photon.


In \gls*{NLO} calculations within the framework of perturbative quantum field theory, the amplitude for a given process includes contributions from $n$-particle states, comprising both the Born contribution and the virtual corrections, as well as $n+1$-particle states due to real emission processes. The POWHEG method \cite{Frixione:2007vw}, an established technique for matching \gls*{NLO} calculations with \gls*{PS}, effectively integrates the first radiation emission into the $n$-particle state. This integration is achieved by employing a modified Sudakov form factor, $\Delta_R(p_T)$, which is defined as
\begin{equation}
    \Delta_R(p_T) \sim \exp \left[ - \int \dd \Phi_R \frac{R(\Phi_B,\Phi_R)}{B(\Phi_B) }\theta( k_T(\Phi_B, \Phi_R) - p_T)\right].
    \label{eq:sudakov}
\end{equation}
Here, the integration spans the real emission $n+1$ phase space $\Phi_R$, and the ratio of the real emission correction to the Born cross section, $R(\Phi_B,\Phi_R)/B(\Phi_B)$, stands in for the emission probability, modulated by a Heaviside step function $\theta$ prohibiting emissions below the transverse momentum scale $p_T$. Intuitively, the Sudakov form factor describes the probability of no emission. This interpretation can be verified by inspecting the phase space region where the real corrections diverges ($R\to \infty$) corresponding to a large probability of emission and a vanishing Sudakov factor. To maintain \gls*{NLO} accuracy with this modified factor, one must ensure that the subsequent \gls*{PS} generates radiation only below the established scale $p_T$. In the Les Houches Event (LHE) standard \cite{Alwall:2006yp} this scale of the hard process is defined as the SCALUP parameter and the \gls*{PS} uses it as the starting point to attach softer emissions \cite{Bierlich:2022pfr}. In a $p_T$-ordered parton shower, this is achieved by starting the shower evolution at a given scale, while in an angular-ordered \gls*{PS} a veto algorithm is employed \cite{Nason:2004rx}.

In POWHEG, the process of attaching an emission to Born-like configurations is managed through a probabilistic method involving the transverse momentum of the emission. The procedure starts with the determination of $p_T$ via a randomly selected value $r \in (0,1)$, which is used to compute
\begin{equation}
    \label{eq:hitmiss}
    \log \Delta^{U}(p_T) = \log(r)\,,
\end{equation}
where $\Delta^{U}(p_T)$ represents the lower bound of the Sudakov form factor from \Cref{eq:sudakov}. The lower bound is obtained by replacing the ratio of real emission and Born cross sections $f \sim \frac{R}{B}$ by an upper bounding function $U$. The emission at a given $p_T$ is then subjected to an acceptance probability $P_{\text{acc}} = \frac{f}{U}$. Should the emission not meet the acceptance criteria, it is rejected with a probability of $1 - \frac{f}{U}$, and the procedure is repeated for a decreased $p_T$ value. This iterative process continues until either the emission is accepted or the threshold is reached.

\begin{figure}
\begin{align*}
        \raisebox{0.9cm}{$2 \to 3$:}&
\underbrace{
        \includegraphics[height=3.5cm]{{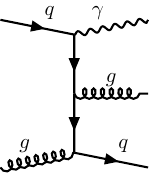}}
}_\text{Compton}
    \quad
\underbrace{
        \includegraphics[height=3.5cm]{{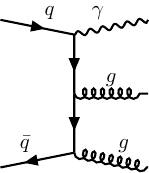}}
}_\text{Annihilation}
    \quad
        \includegraphics[height=3.5cm]{{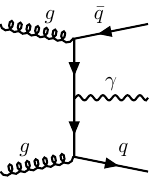}}
    \quad
        \includegraphics[height=3.5cm]{{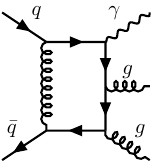}}
\\[-15pt]
        \hphantom{2 \to 3:}&
\underbrace{\hphantom{
        \includegraphics[height=3.5cm]{{img/compton.pdf}}
    \quad
        \includegraphics[height=3.5cm]{{img/annihilation.pdf}}
    \quad
        \includegraphics[height=3.5cm]{{img/gg.pdf}}
}}_\text{Leading Order}
    \quad
\underbrace{\hphantom{
        \includegraphics[height=3.5cm]{{img/virt.pdf}}
}}_\text{Virtual}
    \\
        \raisebox{0.9cm}{$2 \to 4$+:}&
    \quad
        \includegraphics[height=3.5cm]{{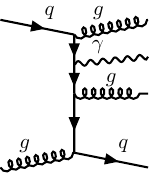}}
    \quad
\underbrace{
        \includegraphics[height=3.5cm]{{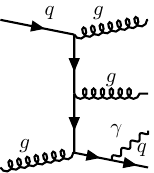}}
    \quad
        \includegraphics[height=3.5cm]{{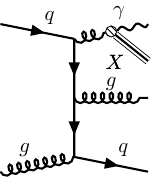}}
}_\text{Fragmentation}
\\[-15pt]
        \hphantom{2 \to 4+:}&
\underbrace{\hphantom{
        \includegraphics[height=3.5cm]{{img/real.pdf}}
    \quad
        \includegraphics[height=3.5cm]{{img/bremsstrahlung.pdf}}
}}_\text{Real}
    \quad
\underbrace{\hphantom{
        \includegraphics[height=3.5cm]{{img/fragmentation.pdf}}
}}_\text{Pythia/Herwig}
\end{align*}
\caption{
    Schematic representation of the different contributions to prompt photon production in association with two jets at NLO+PS.
    This extended disambiguation is inspired by Ref.\ \cite{CERN-LHCC-2020-009}.
}
\label{fig:feyn}
\end{figure}

In \Cref{fig:feyn}, exemplary Feynman diagrams for direct photon production with at least two jets are shown. The addition of a jet to the direct photon process, where the prompt photons come from either Compton or annihilation processes and fragmentation photons from bremsstrahlung or gluon/quark fragmentation, is straightforward. The main conceptual difference to the direct photon production with one jet is that already at leading order there is a gluon-gluon initiated process. As can be seen from the figure, going beyond leading order the definition of fragmentation photons becomes ambiguous, as the splitting $q\to q\gamma$ can also be attributed to the real correction.

The \gls*{POWHEG} method in combination with a \gls*{PS} can be considered an alternative description to the fragmentation process. In the following, we will explain the main idea using the example of the simplest process $pp \to \gamma j$ at NLO, which straightforwardly extends to $pp \to \gamma jj$. The \gls*{NLO} real photon production at $\mathcal{O}(\alpha \alpha_S^2)$ in \gls*{POWHEG} comprises two primary contributions: events with an underlying photon and a parton, and events with two underlying partons. The first category represents the true direct photon contribution, which includes the $q\bar{q} \to \gamma g$ (annihilation) or $g q \to \gamma q$ (Compton scattering) processes. The second category involves the dijet process, which contributes to the fragmentation component. NLO events are generated from these two contributions in the following ways. For the direct photon contribution, POWHEG may attach a gluon, setting the SCALUP parameter to the transverse momentum of the gluon. Alternatively, it may split a gluon into a quark-antiquark pair, with the SCALUP set to the $p_T$ of one of the quarks, or attach nothing, setting the SCALUP to a value of perturbative cutoff at approximately 1 GeV. It is important to note that $\gamma \to q \bar{q}$ splittings remove photons from the final state, thus affecting observables that tag photons only at higher orders, particularly if the photon is regenerated through \gls*{QED} showers. For the dijet contribution, POWHEG may attach a photon, setting the SCALUP parameter to the $p_T$ of the photon. \gls*{QCD} radiation on top of dijet events, which do not initially include photons, contributes at higher orders if the photon is regenerated through \gls*{QED} showers. We do not include such \gls*{QCD} real corrections to the underlying \gls*{QCD} Born events.
The inclusion of underlying dijet events fulfils the purpose of subtracting the collinear singularity and is not intended for a fragmentation function-like convolution.
The remainder of the fragmentation contributions is included by the \gls*{PS}.
	\begin{table}
		\centering
		\begin{tabular}{|c|c|c|c|c|}
			\hline
			Born \vphantom{\Big(} & \multicolumn{2}{c|}{\cellcolor[gray]{.8}$pp\to \gamma jj\sim\mathcal O(\alpha \alpha_s^2)$} & \multicolumn{2}{c|}{\cellcolor[gray]{.8}$pp\to jjj\sim\mathcal O(\alpha_s^3)$} \\
			\hline
			Virtual \vphantom{\Big(} &$\mathcal O(\alpha \alpha_s^2)\cdot\mathcal O(\alpha)$  &\cellcolor[gray]{.8}$\mathcal O(\alpha \alpha_s^2)\cdot\mathcal O(\alpha_s)$ & $\mathcal O(\alpha_s^3)\cdot\mathcal O(\alpha)$  &$\mathcal O(\alpha_s^3)\cdot\mathcal O(\alpha_s)$ \\
			\hline
			Real \vphantom{\Big(} & $pp\to \gamma\gamma jj\sim\mathcal O(\alpha^2 \alpha_s^2)$ & \multicolumn{2}{c|}{\cellcolor[gray]{.8}$pp\to \gamma jjj \sim \mathcal O(\alpha \alpha_s^3)$} & $pp\to jjjj \sim \mathcal O(\alpha_s^4)$\\
			\hline
		\end{tabular}
        \caption{
            Processes contributing to prompt photon production in association with two jets at NLO.
            Only the processes highlighted in grey are included in the calculation.
        }
        \label{tab:proc}
	\end{table}
\Cref{tab:proc} orders prompt photon production by the couplings and shows which processes are included in the calculation highlighted in grey.

The above treatment has a caveat. As attaching a photon to a parton is suppressed by the electromagnetic coupling constant $\alpha$, it will only happen at a very small rate. Consequently, most of the events generated will be dijet-like events, with few events containing photons in the final state. Therefore, in Ref.~\cite{Jezo:2016ypn} the \enquote{enhancedrad} parameter $c$ was introduced to increase the ratio of generated events with photons. As the same problem remains for the process with one extra jet, we repeat the same procedure again in analogy to \cite{Hoeche:2009xc} and similar to \texttt{PYTHIA}'s \enquote{Enhance}-options \cite{Lonnblad:2012hz}. The probability of attaching a photon is increased by modifying \Cref{eq:hitmiss} to
\begin{align}
        \log \Delta^{cU}(p_T) = \log(r) &&\Leftrightarrow &&\log \Delta^{U}(p_T) = \frac{\log(r)}{c}\,,
\end{align}
where a constant $c$, $c>1$, is introduced to increase the probability of attaching a photon. Finally, the event weight has to be adjusted for all previous $n$ rejected emissions,
\begin{equation}
            w_n = \frac 1 c \prod_{i=1}^n \frac{1- \frac{f_i}{cU_i}}{1-\frac{f_i}{U_i}}
            \,.
            \label{eq:weight}
\end{equation}
This allows us to increase the ratio of photons in typical events from $\approx \SI{5}{\percent}$ at $c=1$ to $\approx 30-40\si{\percent}$ at $c=10$ up to $\SI{90}{\percent}$ at $c=100$ in exchange for reduced weights.

\subsection{Calculation and implementation in POWHEG\,BOX\,V2}
\label{sec:nlo}

The numerical implementation of the \gls*{NLO} calculation for the process $pp \to \gamma jj$ was performed using the {\tt POWHEG\,BOX\,V2} framework \cite{Alioli:2010xd}. Usually, the required ingredients for a new process are the Born amplitudes and their colour- and spin-correlated counterparts, the Born phase space, a decomposition of the amplitudes in the colour flow basis, the finite part of the virtual corrections, and the real correction amplitudes. For this process, however, we will also reuse the modifications, e.g.~the Sudakov weight \Cref{eq:weight}, to the core {\tt POWHEG\,BOX} functionality introduced in \cite{Jezo:2016ypn}.

The usual leading order, virtual and real corrections were computed with \texttt{FormCalc} \cite{Hahn:1998yk} in a five-flavour scheme. The required leading order components for  $pp \to jjj$ have been reused from the trijet process \cite{Kardos:2014dua}. The ultraviolet divergences are handled by the constrained differential renormalisation procdure \cite{delAguila:1998nd}. Infrared divergences are treated using the \gls*{FKS} subtraction method \cite{Frixione:1995ms,Frixione:1997np} in \texttt{POWHEG\,BOX\,V2}, which combines the colour- and spin- correlated amplitudes to reproduce and identify the divergent behaviour. The colour-correlated Born amplitudes were obtained from \texttt{FormCalc} by injecting the colour charge operators into the amplitudes in Mathematica,
\begin{equation}
    B_{ij} = -N \sum_{\substack{\text{spins}\\\text{colours}}} \mathcal{M}_{\{c_k\}} \mathcal{T}^a_{c_i,c_i'} \mathcal{T}^a_{c_j,c_j'} [\mathcal{M}_{\{c_k\}}^\dagger]_ {\substack{c_i\to c_i' \\ c_j\to c_j'}}
    \,.
\end{equation}
Here the averaging factors are included in $N$, $\{c_k\}$ are the sets of colour indices of the external particles, and $\mathcal{T}$ represents the structure constants $f^{abc}$ for gluons or $\pm T^a_{\alpha\beta}$ for quarks. The colour correlation is automatically checked at runtime by requiring that $\sum_{i,i\neq j} B_{ij} = C_{f_j} B$ where $C_{f_j}$ is the Casimir constant of the parton $j$. Similarly to the colour correlations, in the expression for the spin correlation
\begin{equation}
    B^{\mu\nu}_j = N \sum_{\substack{\text{spins}\\\text{colors}}} \mathcal{M}_{\{s_k\}} \epsilon^{\mu*}_{s_j} \epsilon^\nu_{s_j'} [\mathcal{M}_{\{s_k\}}^\dagger]_ {\substack{s_j\to s_j'}}
\end{equation}
the identity between spins $s_j$ and $s_j'$ has been replaced by the polarization vectors $\epsilon^\mu_{s_j}$. As \texttt{FormCalc}'s exported Fortran code uses the spinor helicity formalism \cite{Dixon:2013uaa} to compute the matrix elements, it is possible to identify individual polarised and averaged helicity amplitudes. Thus one obtains the spin correlations between the photon and the partons by taking all but one external particle as unpolarised. The four combinations of $s_j \cross s_j'$ are scaled by the corresponding numerical values of the polarisation vectors and then summed. A trivial cross check is to verify $g_{\mu\nu} B_j^{\mu\nu} = -B$ for all $j$, which {\tt POWHEG\,BOX\,V2} also does at runtime. The inclusion of colour flow information is necessary to enable the parton shower to properly manage colour correlations beyond the hard process. We perform the colour flow decomposition analogously to Refs.~\cite{Maltoni:2002mq,BENGTSSON1984323,Kilian:2012pz}.

In the course of our study, several computational tools and frameworks were employed to ensure the reliability and accuracy of our results in prompt photon production calculations. To verify the computational precision of our \gls*{NLO} \gls*{QCD} calculations, we compared to multiple automatic matrix element generators. The virtual corrections were validated against \texttt{MadGraph5\_aMC@NLO}~\cite{Alwall:2014hca}, \texttt{OpenLoops2}~\cite{Buccioni:2019sur} and \texttt{Recola\,2}~\cite{Denner:2017wsf}, whereas the real corrections were checked with \texttt{FormCalc} and \texttt{OpenLoops2}. This comprehensive cross-check confirms the consistency across different tools. After evaluating the performance and numerical quality of the \gls*{FKS} subtraction of the various computations, \texttt{OpenLoops2} was selected for the final implementation and numerical computations in the remainder of this study. During the process of comparing different calculations, we also developed an interface between \texttt{MCFM} \cite{Campbell:2019dru} and \texttt{Rivet}/\texttt{YODA}\cite{Bierlich:2019rhm,Buckley:2023xqh}, which also allows for direct comparisons with experimental analyses (cf.~appendix \ref{app:mcfm}).

For the definition of the phase space, we adopted the multi-channel phase space construction from the topologically similar trijet {\tt POWHEG\,BOX\,V2} process~\cite{Kardos:2014dua}. This method provides independent importance sampling in the six divergent final-state regions (FSR) and three initial-state regions (ISR). They diverge as
\begin{align}
    S_{0j}^\text{ISR} &=S_{1j} + S_{2j} = \frac{1}{E_j^2 (1-\cos^2\theta_{1j})} && \text{and} && S_{ij}^\text{FSR} = \frac{E_i^2+E_j^2}{2 E_i^2 E_j^2 (1-\cos\theta_{ij})}
\end{align}
with $i,j \geq 3$. After normalisation, they read
\begin{align}
    \tilde S_{0j}=\frac{S_{ij}}{\sum_{j} (S_{0j} + \sum_{i}S_{ij} ) },&&
    \tilde S_{ij}=\frac{S_{ij}}{\sum_{j} (S_{0j} + \sum_{i}S_{ij} ) } \frac{E_j}{E_i+E_j}\,,
\end{align}
where the ratio over the energies is introduced to raise the degeneracy between $i$ and $j$. Starting from a $2\to2$ massless phase space ($\Phi_{2\to2}$), the three-particle phase space is constructed as $\dd{\Phi_B} = \sum_{kj}\tilde S_{kj}\dd \Phi_{2\to3,kj}$, where $\dd \Phi_{2\to3,kj}$ is $\dd\Phi_{2\to2}$ with the emission $kj$ added through {\tt POWHEG\,BOX}'s $n+1$ phase space construction routines \cite{Nason:2004rx,Frixione:2007vw,Alioli:2010xd}.
If one now probes a divergent $2\to3$ regime, then the non-divergent $\tilde S$ vanish by construction
\begin{align}
\tilde S_{0j,ij}\to \begin{cases}
    1 & \text{as } E_j \to 0 \text{ or }  \theta_{ij} \to 0 \,, \\
    0 & \text{else}\,.
    \end{cases}
\end{align}
Thus, the phase space sampling is guaranteed to pick a $\dd \Phi_{2\to3,kj}$ that works well despite the soft and/or collinear divergences.

\section{Isolated prompt photon production with two jets at \boldmath$\sqrt{s}=13$ TeV}
\label{sec:res}

\subsection{Simulation setup}
\label{sec:simulation}

We set the renormalisation and factorisation scales equal to the transverse momentum of the photon, $E_T^\gamma=p_T^\gamma$. The PDF in use is MSHT20nlo \cite{Bailey:2020ooq} provided through \texttt{LHAPDF6} \cite{Buckley:2014ana}, and the running of $\alpha_S$ is calculated internally by \texttt{POWHEG\,BOX}. In all of the following figures in \Cref{sec:atlas} and \Cref{sec:iso}, the vertical error bars represent the statistical uncertainties, while the bands represent the scale uncertainties. The scale uncertainties are estimated by the standard factor-two seven-point variation of the renormalisation and factorisation scales by a factor of two around the central value while excluding relative factors of four. As the computations involves both tree level diagrams of $pp \to \gamma j j j$ and $pp \to j j j$ this takes a significant amount of CPU hours. In order to speed up the process, we only enabled the virtual corrections in the final reweighting step, instead of generating equal weighted events for all processes.\footnote{Due to the inclusion of the Sudakov reweighting factor \Cref{eq:weight}, the events are already not equally weighted, thus this is not significant loss.}

The events from \texttt{POWHEG\,BOX\,V2} in the LHE format are showered both by \texttt{Pythia8} or \texttt{Herwig7}, and then analysed by \texttt{Rivet3}~\cite{Bierlich:2019rhm}. We have made significant updates to our {\tt Pythia8} interface\footnote{The interface will be made publicly available in the near future.}~\cite{Jezo:2018yaf,FerrarioRavasio:2023jck} to enhance its compatibility with the latest of tools it relies on~\cite{Dobbs:2001ck,Sjostrand:2006za,Sjostrand:2014zea,Bierlich:2019rhm,Verbytskyi:2020sus}. Additionally, we have expanded its flexibility by exposing several meta parameters, such as toggles for \gls*{QCD} and \gls*{QED} showers, to the users through the \verb|powheg.input| file (for an example see \cref{lst:input} in \cref{app:input}). This advancement simplifies usage of the program by providing options that were previously accessible only through customised settings. Mirroring this enhanced configurability, we have also developed a new interface for \texttt{Herwig7}~\cite{Bahr:2008pv,Bellm:2015jjp,Bewick:2023tfi}. It incorporates similar user-exposed options, thereby standardising the setup across different parton shower generators and making it easier for users to transition between tools without reconfiguring their parameters extensively. The complexity of keeping \texttt{Pythia8} and \texttt{Herwig7} aligned can be seen from \cref{lst:pythia,lst:herwig} in \cref{app:shower}.

\subsection{Comparison with ATLAS data}
\label{sec:atlas}

The following figures show the results of our new calculation for the production of an isolated photon plus two jets in the context of a measurement performed by ATLAS in proton-proton collisions at $\sqrt{s}=13$ TeV~\cite{ATLAS:2019iaa}. Besides the cuts defined by the ATLAS detector acceptance, the analysis uses an isolation cone of $R_I = 0.4$ and isolation energy of $E_T^\text{iso} = \SI{10}{GeV} + 0.0042 p_T^\gamma + \pi R_I^2 \rho^{p_T}_j(\eta)$. The jet density $\rho^{p_T}_j(\eta)$ describes the hadronic activity by averaging the transverse momenta of the jets for a given rapidity $\eta$. Consequently, the photon can be less isolated if it has a large $p_T$ or if there are many jets. After jets are identified through the anti-$k_T$ algorithm with a jet radius $R_J=0.4$, they are sorted by transverse momentum. That is the leading jet ($\text{jet}_1$) is the one with the highest transverse momentum, the subleading jet ($\text{jet}_2$) is the one with the second highest, etc. The same nomenclature would apply for photons, but here we are only interested in the hardest photon. The analysis examines several observables in various regimes, namely inclusive, direct-enriched and fragmentation-enriched contributions. The corresponding Rivet implementation is \href{https://rivet.hepforge.org/analyses/ATLAS_2019_I1772071.html}{ATLAS\_2019\_I1772071}.
In addition to the experimental data, \gls*{NLO} \gls*{QCD} predictions from \texttt{Sherpa}~\cite{ATLAS:2019iaa}, our new calculation of direct photons in \texttt{POWHEG} with \texttt{PYTHIA8} (\texttt{PY8}) and \texttt{HERWIG7} (\texttt{HW7}) as well as recent \gls*{NLO} and \gls*{NNLO} calculations~\cite{Badger:2023mgf} are displayed. The \texttt{Sherpa} calculation merges $\gamma+1j$ and  $\gamma+2j$ NLO with $\gamma + 3j$ and $\gamma+4j$ LO events and supplements it with a parton shower \cite{Krauss:2001iv,Cascioli:2011va,Schumann:2007mg,Hoeche:2012yf}.
The \gls*{NLO} and \gls*{NNLO} predictions do not include a fragmentation contribution, arguing that it should not exceed \SI{5}{\percent} for the inclusive and direct-enriched regime. Both the (N)NLO and the NLO merged calculations use a smooth cone photon isolation at the matrix-element level. While this publication includes only figures with photon observables from the ATLAS analysis, the complete set of distributions is available in the ancillary files accompanying this publication.

\begin{figure}
    \centering
    \begin{subfigure}[t]{0.45\textwidth}
        \centering
        \includegraphics[width=6.5cm]{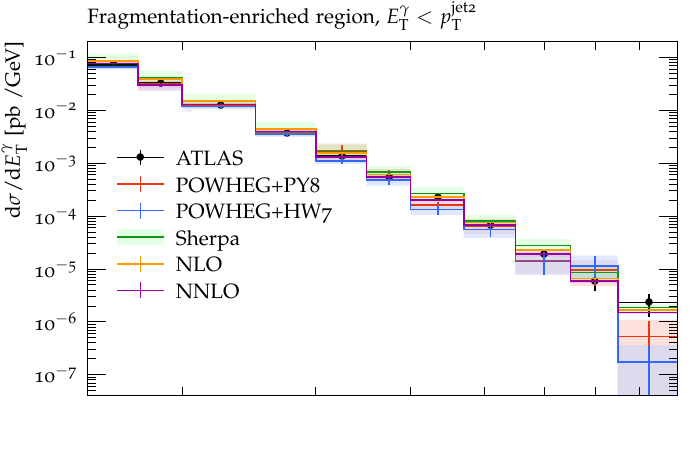}
        \vspace{-2.30em}

        \includegraphics[width=6.5cm]{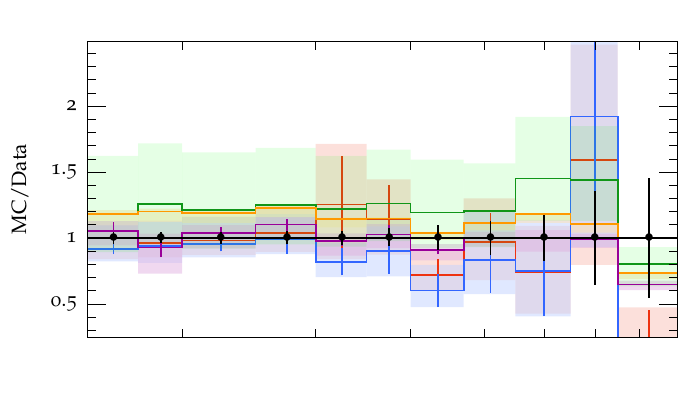}
        \vspace{-2.30em}

        \includegraphics[width=6.5cm]{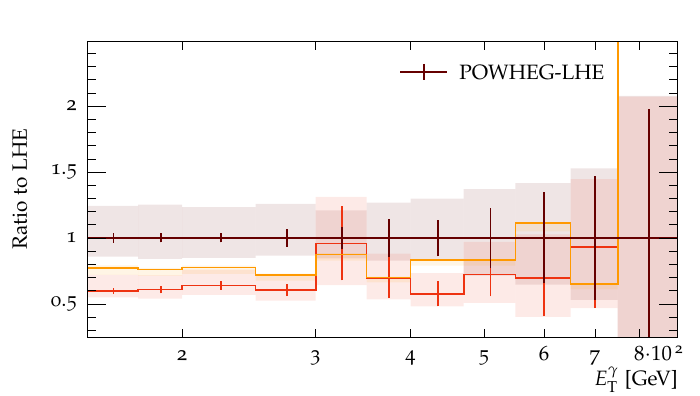}
        \vspace{-0.00em}

        \caption{Photon transverse momentum}
        \label{fig:frag:pt}
    \end{subfigure}
    \hfill
    \begin{subfigure}[t]{0.45\textwidth}
        \centering
        \includegraphics[width=6.5cm]{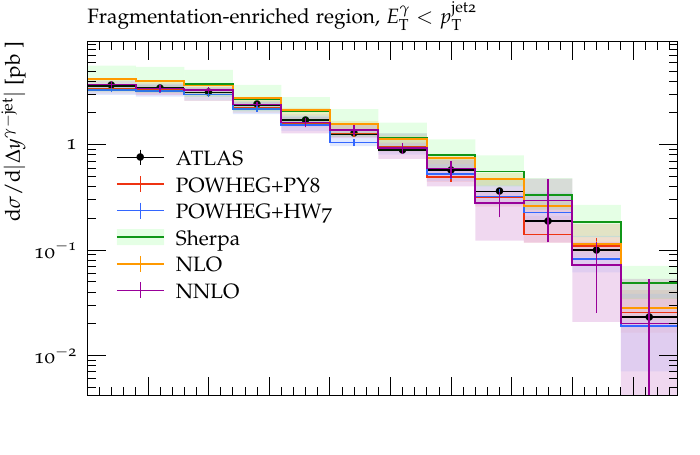}
        \vspace{-2.30em}

        \includegraphics[width=6.5cm]{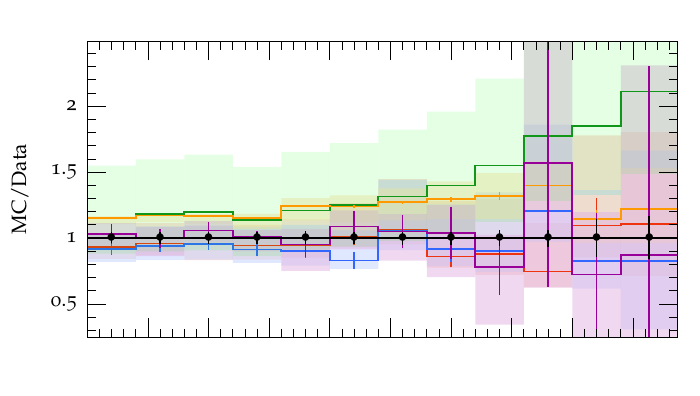}
        \vspace{-2.30em}

        \includegraphics[width=6.5cm]{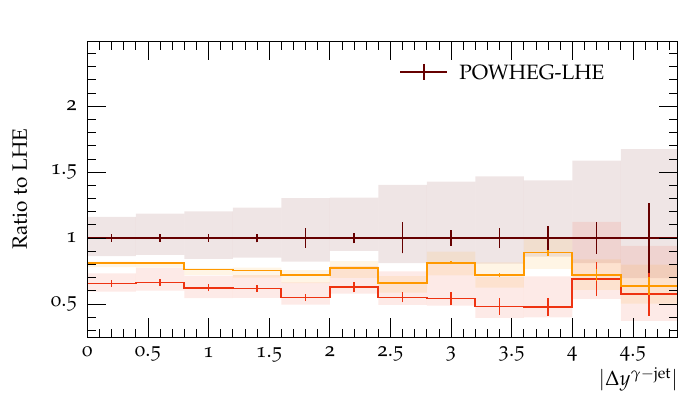}
        \vspace{-0.00em}

        \caption{Rapidity difference between the hardest photon and each of the two hardest jets}
        \label{fig:frag:dy}
    \end{subfigure}
    \begin{subfigure}[t]{0.45\textwidth}
        \centering
        \includegraphics[width=6.5cm]{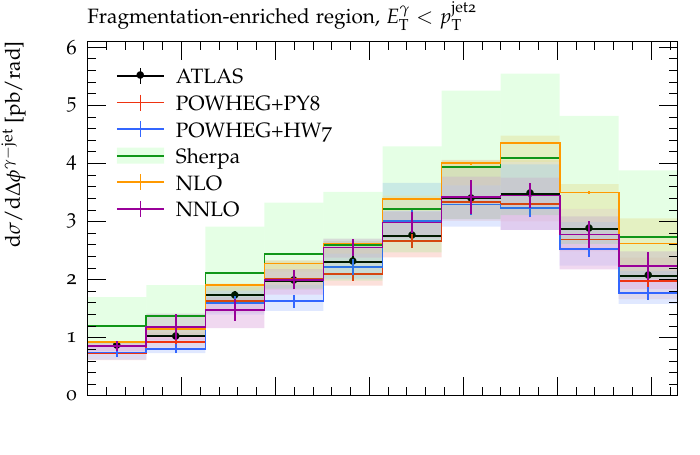}
        \vspace{-2.30em}

        \includegraphics[width=6.5cm]{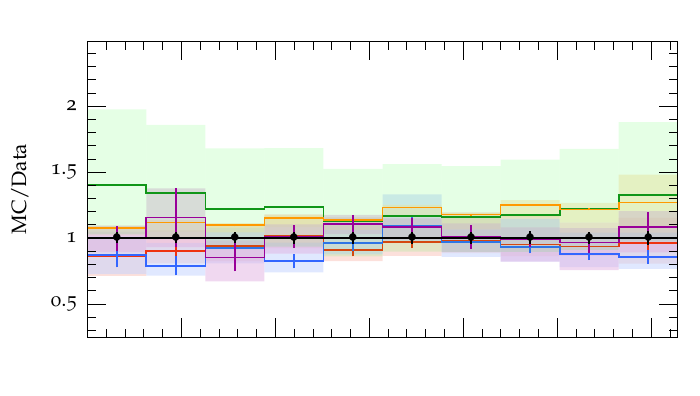}
        \vspace{-2.30em}

        \includegraphics[width=6.5cm]{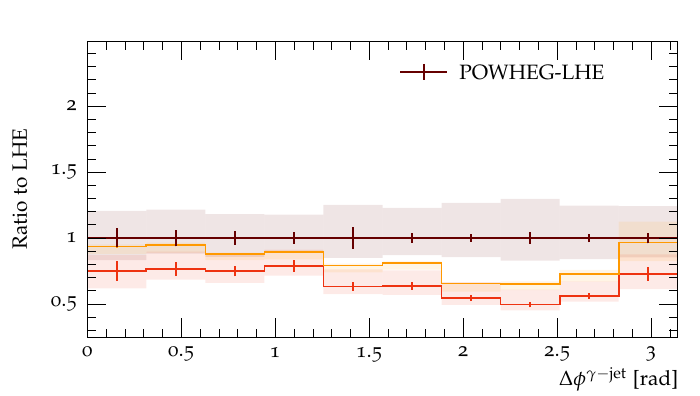}
        \vspace{-0.00em}

        \caption{Azimuthal angle difference between the hardest photon and each of the two hardest jets}
        \label{fig:frag:dph}
    \end{subfigure}
    \hfill
    \begin{subfigure}[t]{0.45\textwidth}
        \centering
        \includegraphics[width=6.5cm]{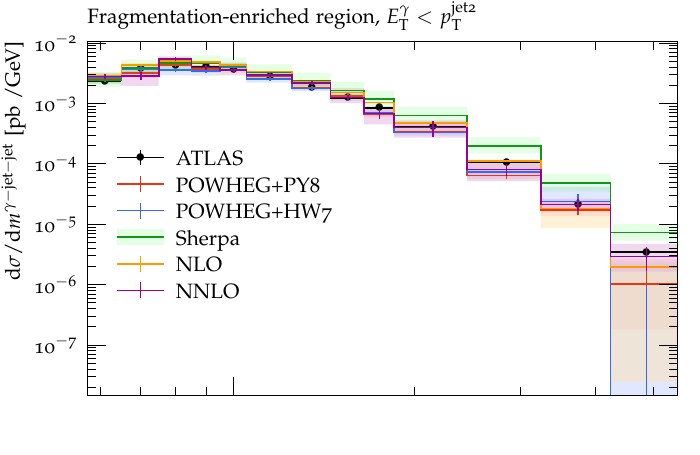}
        \vspace{-2.30em}

        \includegraphics[width=6.5cm]{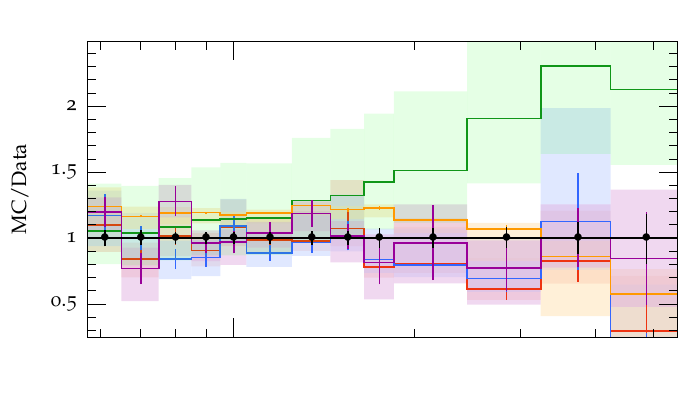}
        \vspace{-2.30em}

        \includegraphics[width=6.5cm]{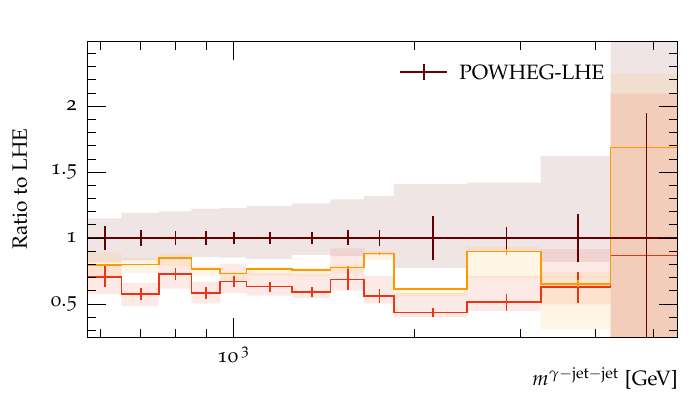}
         \vspace{-0.00em}

       \caption{Invariant mass of the photon-jet-jet system}
        \label{fig:frag:m}
    \end{subfigure}
    \caption{Differential cross sections with respect to different observables in the fragmentation-enhanced regime as defined by ATLAS~\cite{ATLAS:2019iaa}.}
    \label{fig:frag}
\end{figure}

We first inspect the fragmentation-enhanced regime, which is defined by requiring both jets to be harder than the isolated prompt photon, $E_T^\gamma < p_T^{{\rm jet}_2}$. As can be seen in \Cref{fig:frag:pt}, the {\tt POWHEG+PY8} prediction describes the data well. The {\tt HW7} parton shower predicts a slightly lower rate than {\tt PY8}, but still agrees reasonably well. The central prediction from {\tt Sherpa} instead lies appreciably higher than the data, but it still includes the data within its relatively large uncertainty band. The \gls*{NNLO} prediction in the fragmentation-enriched regime was not shown in its publication \cite{Badger:2023mgf} due to the omission of fragmentation contributions, but surprisingly, it still describes the data well despite remaining large numerical and scale uncertainties.
A similar picture persists in \Cref{fig:frag:dph}, where {\tt Sherpa} and the \gls*{NLO} curve are consistently 10-20\% above the data, while \gls*{NNLO} and POWHEG are in agreement with the data and each other. Obtaining adequate statistics in the tails of the distributions of \Cref{fig:frag:dy} and \Cref{fig:frag:m} is challenging. This results in quite some fluctuations in both the NLO+PS and the fixed order \gls*{NNLO} predictions. The prevailing trend also here is that the \gls*{NNLO} and POWHEG predictions, which overlap within uncertainties, describe the data best. We note that the shower corrections beyond NLO are quite large for all observables (10-20\%), see the second ratio panel. The first POWHEG emission correction, already included at the LHE stage, has an appreciable phase space dependence, most notably in the photon-jet angular separation spectrum, and tends to correct in the opposite direction compared to the remaining shower emissions. Nevertheless the total shower correction is relatively flat.

\begin{figure}
    \centering
\begin{subfigure}[t]{0.45\textwidth}
    \centering
    \includegraphics[width=6.5cm]{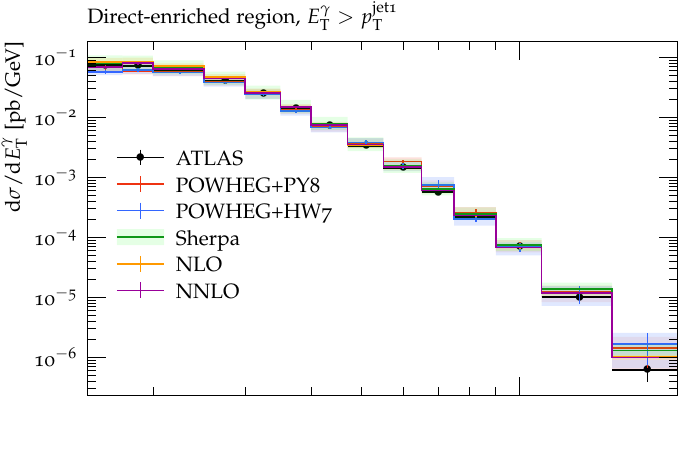}
        \vspace{-2.30em}

        \includegraphics[width=6.5cm]{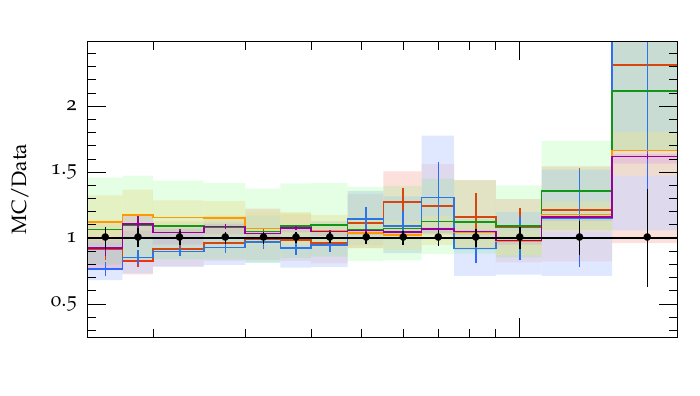}
        \vspace{-2.30em}

        \includegraphics[width=6.5cm]{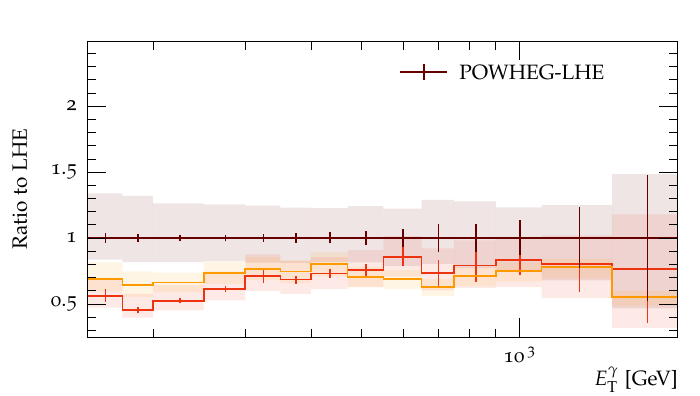}
          \vspace{-0.00em}

  \caption{Photon transverse momentum}
    \label{fig:dir:pt}
\end{subfigure}
\hfill
\begin{subfigure}[t]{0.45\textwidth}
    \centering
    \includegraphics[width=6.5cm]{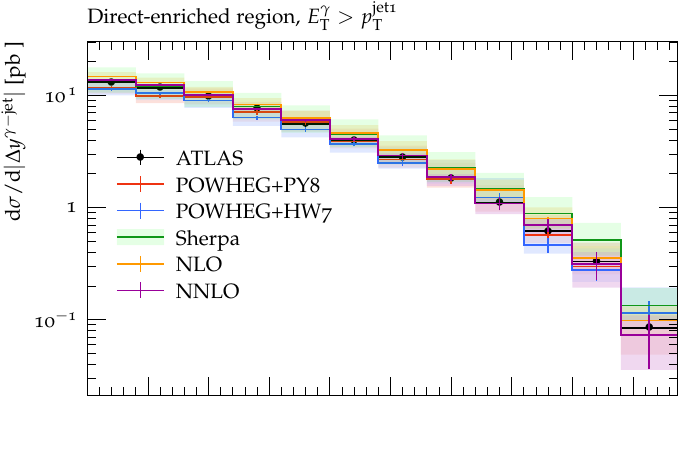}
        \vspace{-2.30em}

        \includegraphics[width=6.5cm]{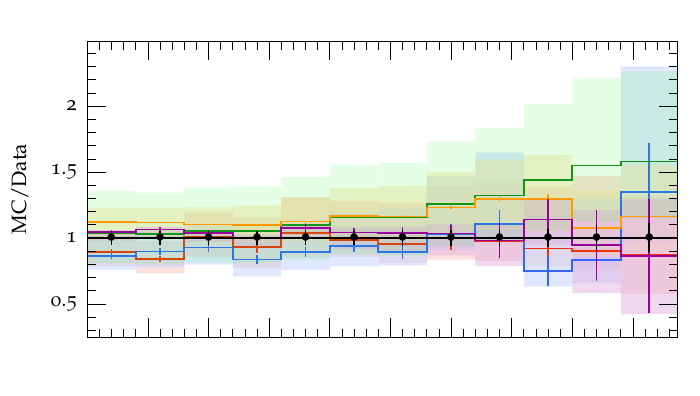}
        \vspace{-2.30em}

        \includegraphics[width=6.5cm]{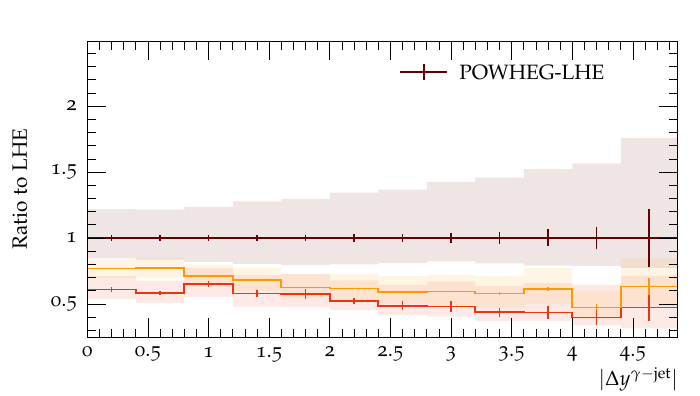}
         \vspace{-0.00em}

   \caption{Rapidity difference between the hardest photon and each of the two hardest jets}
    \label{fig:dir:dy}
\end{subfigure}
\begin{subfigure}[t]{0.45\textwidth}
    \centering
    \includegraphics[width=6.5cm]{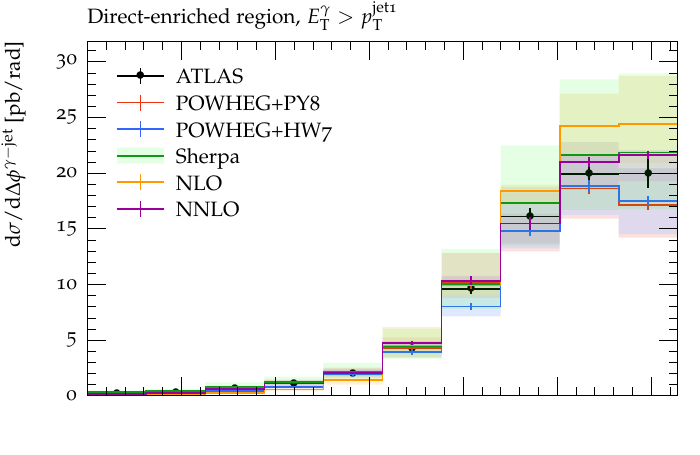}
        \vspace{-2.30em}

        \includegraphics[width=6.5cm]{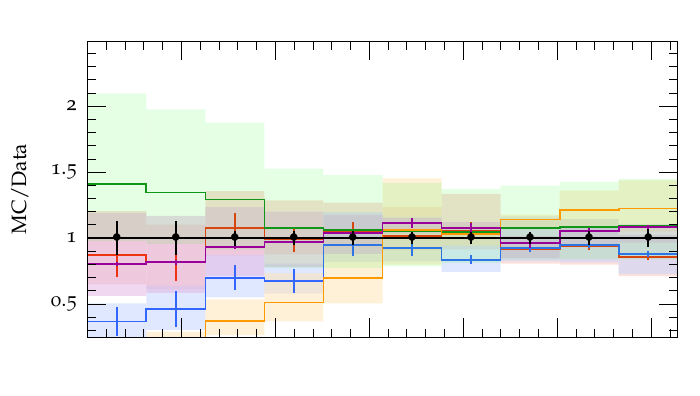}
        \vspace{-2.30em}

        \includegraphics[width=6.5cm]{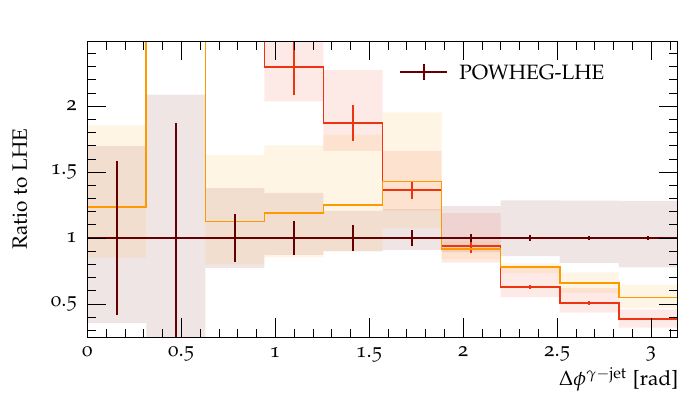}
          \vspace{-0.00em}

  \caption{Azimuthal angle difference between the hardest photon and each of the two hardest jets }
    \label{fig:dir:dph}
\end{subfigure}
\hfill
\begin{subfigure}[t]{0.45\textwidth}
    \centering
    \includegraphics[width=6.5cm]{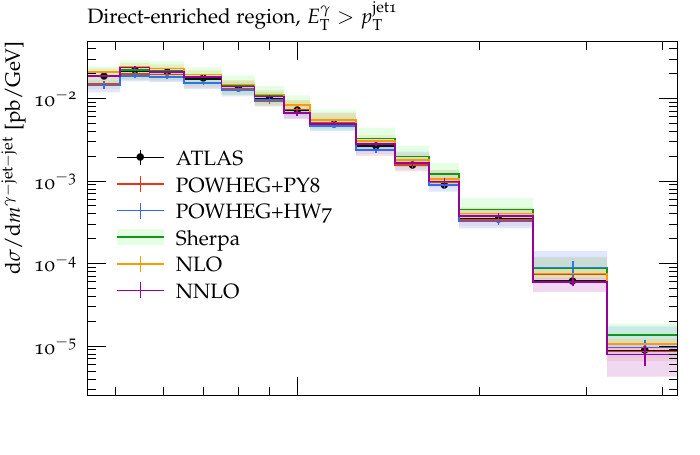}
        \vspace{-2.30em}

        \includegraphics[width=6.5cm]{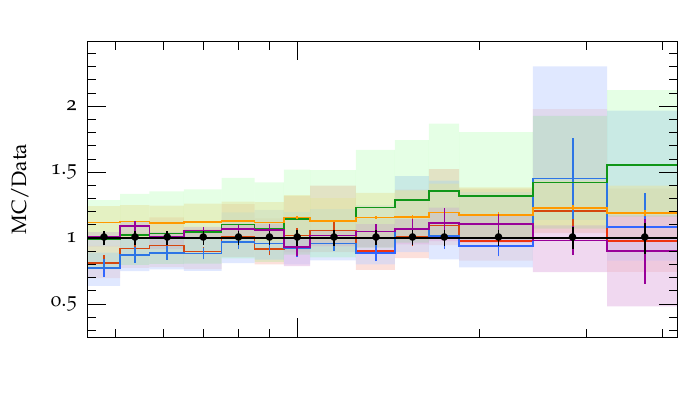}
        \vspace{-2.30em}

        \includegraphics[width=6.5cm]{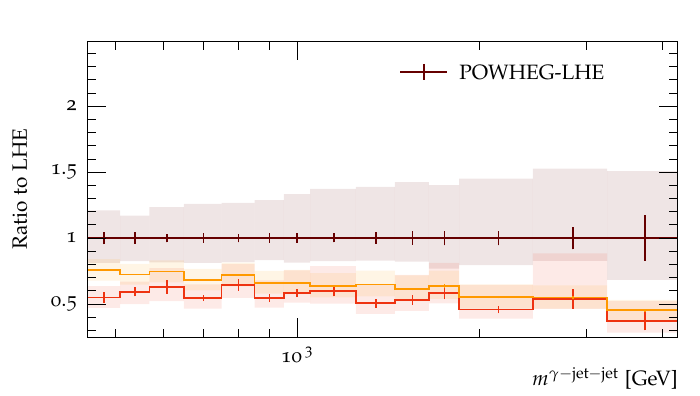}
         \vspace{-0.00em}

   \caption{Invariant mass of the photon-jet-jet system}
    \label{fig:dir:m}
\end{subfigure}
    \caption{Differential cross sections with respect to different observables in the direct-enhanced regime as defined by ATLAS~\cite{ATLAS:2019iaa}.}
    \label{fig:dir}
\end{figure}

In the direct-enhanced regime the photon has a larger transverse momentum than the jets, $E_T^\gamma > p_T^{{\rm jet}_1}$. Thus, this region has enhanced contributions from photons originating from the hard interaction. In comparison to the fragmentation regime, the direct-enhanced regime suffers less from low statistics. The NNLO calculation is in good agreement with the data in all observables, as can be seen in \Cref{fig:dir}. While both {\tt PY8} and {\tt HW7} are mostly close to each other, the {\tt POWHEG+PY8} prediction is often slightly closer to the NNLO curve. The {\tt Sherpa} calculation comes with significantly larger uncertainties than the NNLO or NLO+PS calculations. Furthermore, it seems to overestimate the cross section whenever the cross sections become small. The only very distinct separation between the predictions is in \Cref{fig:dir:dph} at very low azimuthal angle differences between the photon and the jet. This is most likely due to an increased sensitivity to the activity around the photon due to the isolation criterion in this region. There the data points also carry a large uncertainty of about 10\%. While \gls*{NNLO} and {\tt PY8} agree with the data, {\tt HW7} and \gls*{NLO} underestimate them, and {\tt Sherpa} overestimates them by more than 20\%. 
Just as in the fragmentation-enriched region, the shower corrections are relatively large but mostly flat, with the exception of the photon-jet angular separation spectrum where its value is vanishingly small.
\begin{figure}
    \centering
\begin{subfigure}[t]{0.45\textwidth}
    \centering
    \includegraphics[width=6.5cm]{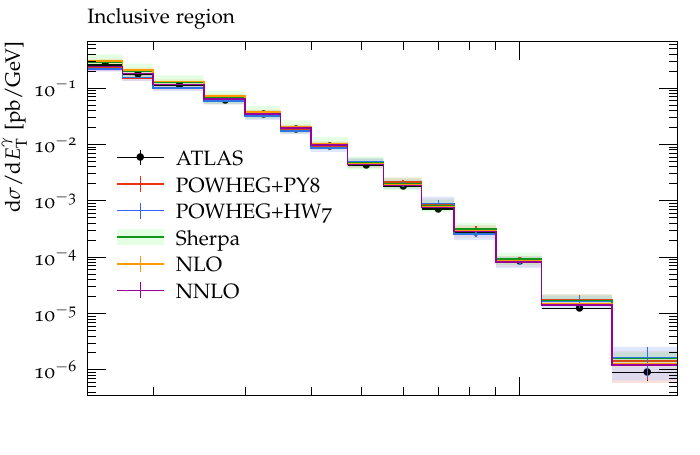}
        \vspace{-2.30em}

        \includegraphics[width=6.5cm]{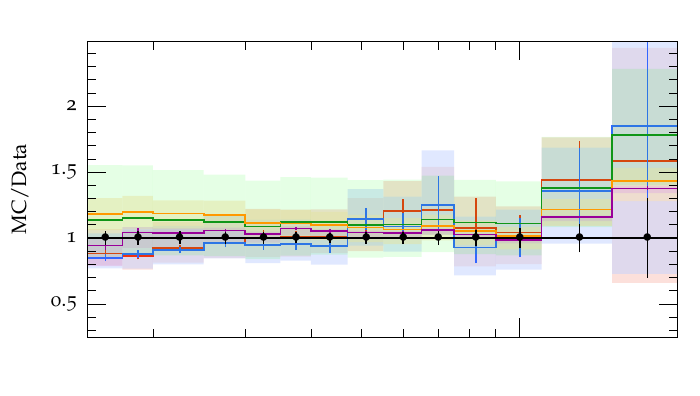}
        \vspace{-2.30em}

        \includegraphics[width=6.5cm]{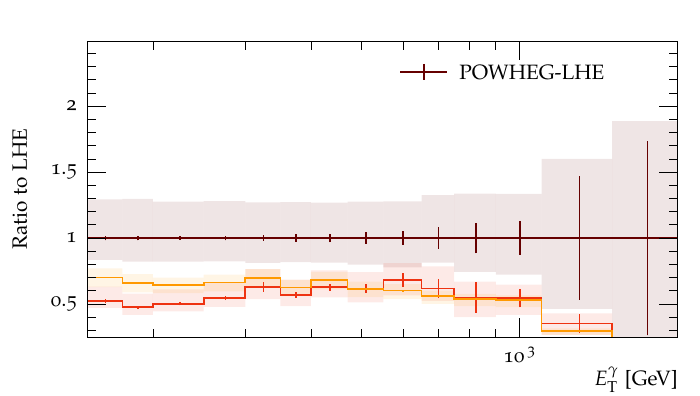}
         \vspace{0.00em}

   \caption{Photon transverse momentum}
    \label{fig:sub1}
\end{subfigure}
\hfill
\begin{subfigure}[t]{0.45\textwidth}
    \centering
    \includegraphics[width=6.5cm]{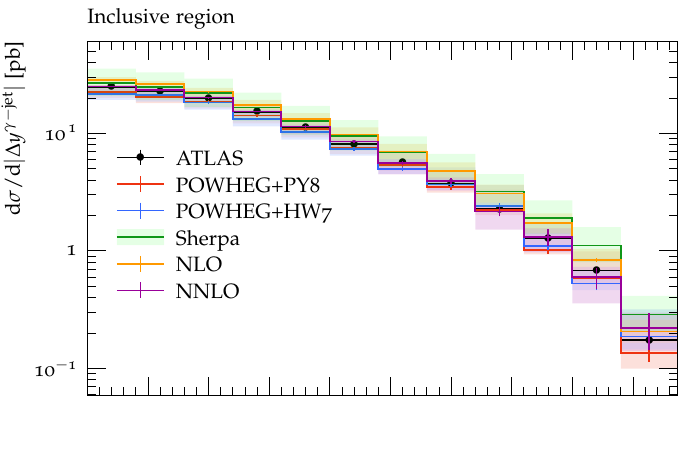}
        \vspace{-2.30em}

        \includegraphics[width=6.5cm]{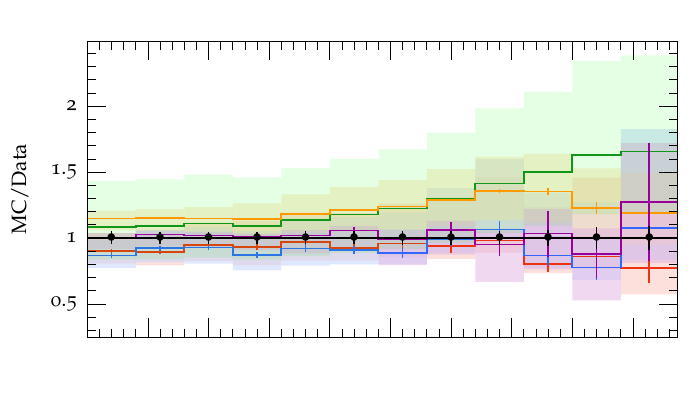}
        \vspace{-2.30em}

        \includegraphics[width=6.5cm]{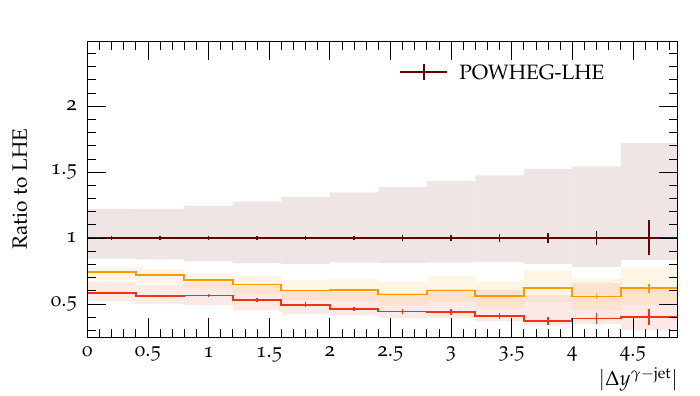}
          \vspace{0.00em}

  \caption{Rapidity difference between the hardest photon and each of the two hardest jets}
    \label{fig:sub2}
\end{subfigure}
\begin{subfigure}[t]{0.45\textwidth}
    \centering
    \includegraphics[width=6.5cm]{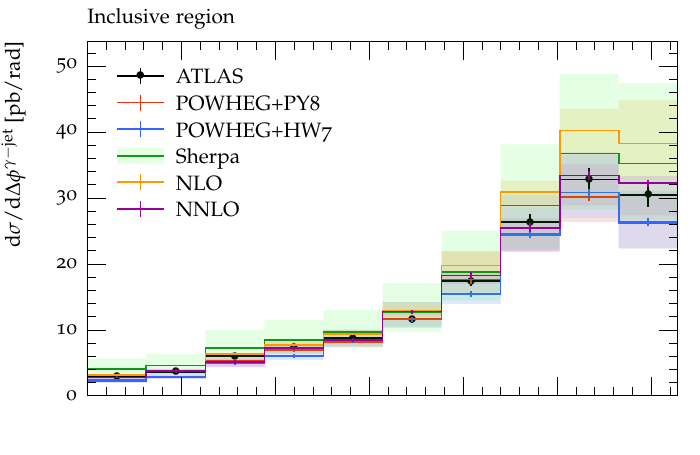}
        \vspace{-2.30em}

        \includegraphics[width=6.5cm]{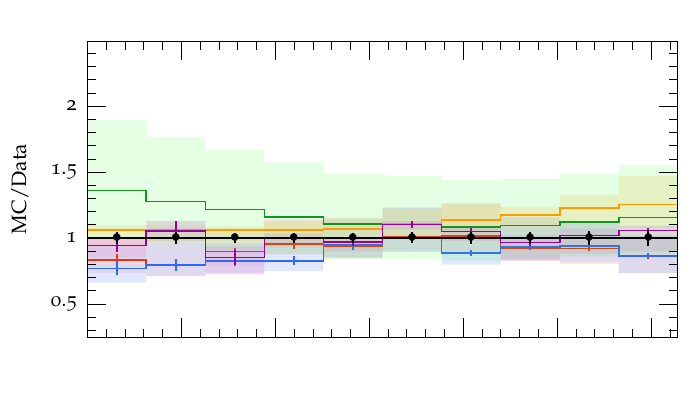}
        \vspace{-2.30em}

        \includegraphics[width=6.5cm]{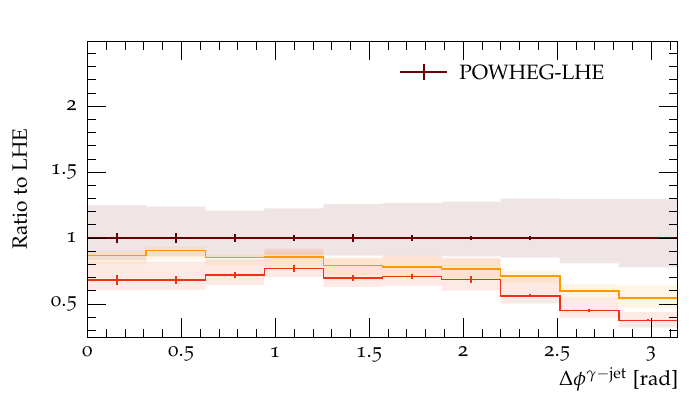}
         \vspace{0.00em}

   \caption{Azimuthal angle difference between the hardest photon and each of the two hardest jets}
    \label{fig:sub3}
\end{subfigure}
\hfill
\begin{subfigure}[t]{0.45\textwidth}
    \centering
    \includegraphics[width=6.5cm]{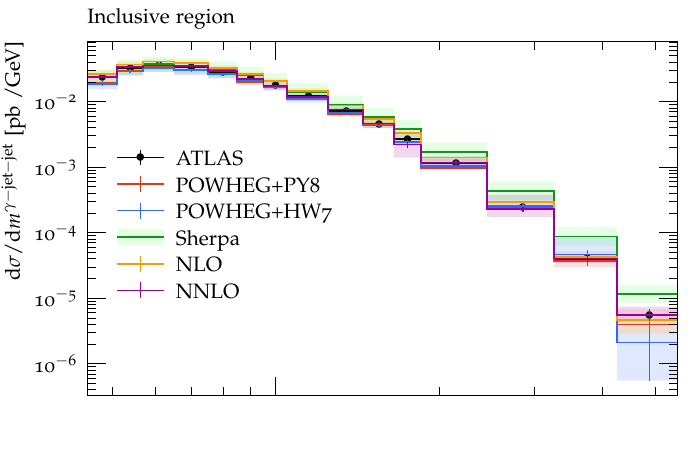}
        \vspace{-2.30em}

        \includegraphics[width=6.5cm]{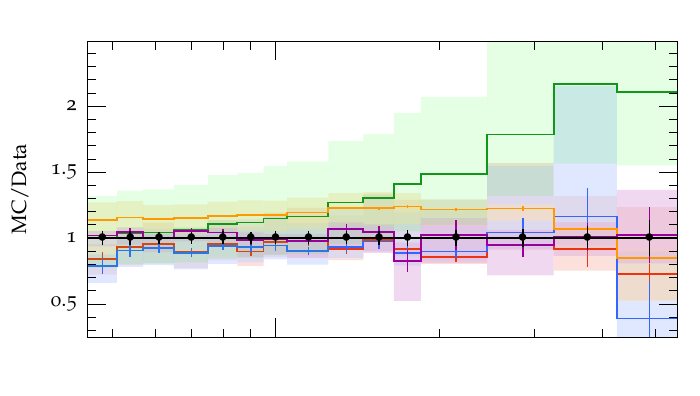}
        \vspace{-2.30em}

        \includegraphics[width=6.5cm]{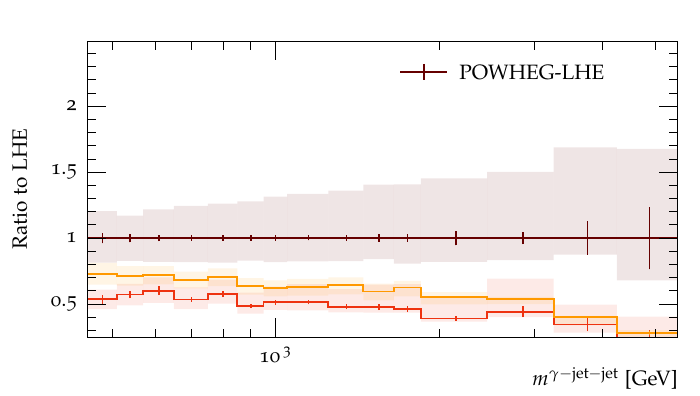}
         \vspace{0.00em}

   \caption{Invariant mass of the photon-jet-jet system}
    \label{fig:sub4}
\end{subfigure}
    \caption{Differential cross sections with respect to different observables in the inclusive regime as defined by ATLAS~\cite{ATLAS:2019iaa}.}
    \label{fig:incl}
\end{figure}

\FloatBarrier{}

\Cref{fig:incl} shows the inclusive predictions, which encompass both of the previous regimes, direct- and fragmentation-enhanced. From comparing \Cref{fig:frag,fig:dir} to \Cref{fig:incl} it is clear that the direct regime contributes more than fragmentation regime, while the remaining contributions can stem from moderately hard photons, $p_T^{{\rm jet}_2} < E_T^\gamma < p_T^{{\rm jet}_1}$. Across all four observables, the \gls*{NNLO} calculation is in best agreement with the data. POWHEG matched either to {\tt PY8} or to {\tt HW7} also describes the data well, but suffers from larger scale and numerical uncertainties than the \gls*{NNLO} calculation. The NLO prediction overestimates the data consistently by about 5-10\%. This leads us to the conclusion that fixed order \gls*{NLO} is not sufficient to describe the data and either \gls*{NNLO} or NLO+PS is required.

Overall the data is described relatively well by all the tools. Both NNLO corrections as well as shower corrections in predictions matched using the POWHEG method, regardless of the shower details, seem to improve the description appreciably. The MC@NLO style matching with the CS dipole based shower from {\tt Sherpa} instead is closer to the fixed order NLO prediction and has been reported with surprisingly large scale uncertainties.

\subsection{Isolation schemes and parton showers}
\label{sec:iso}
\begin{figure}
    \includegraphics[width=0.30\textwidth]{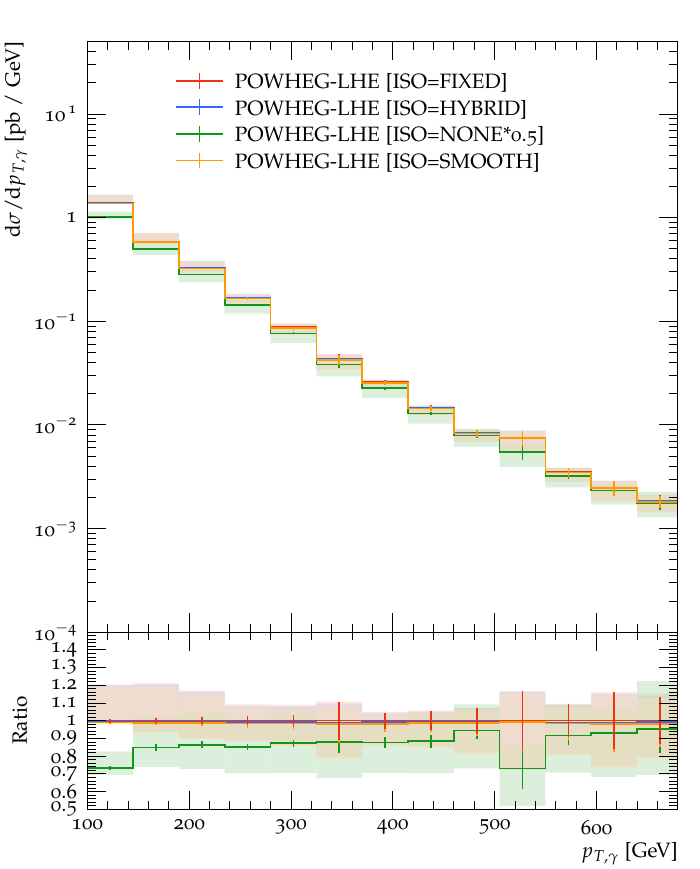}
    \includegraphics[width=0.30\textwidth]{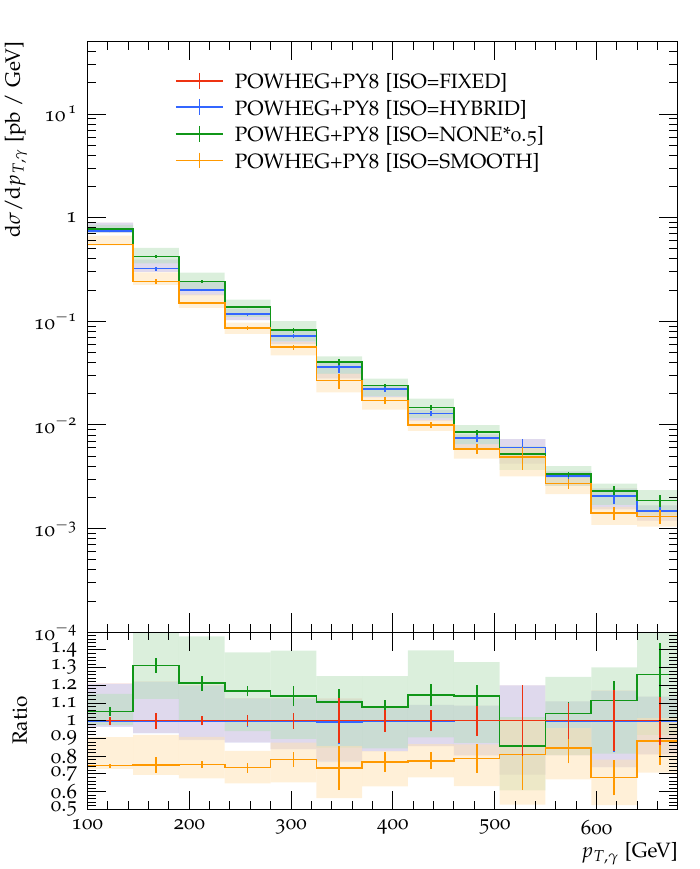}
    \includegraphics[width=0.30\textwidth]{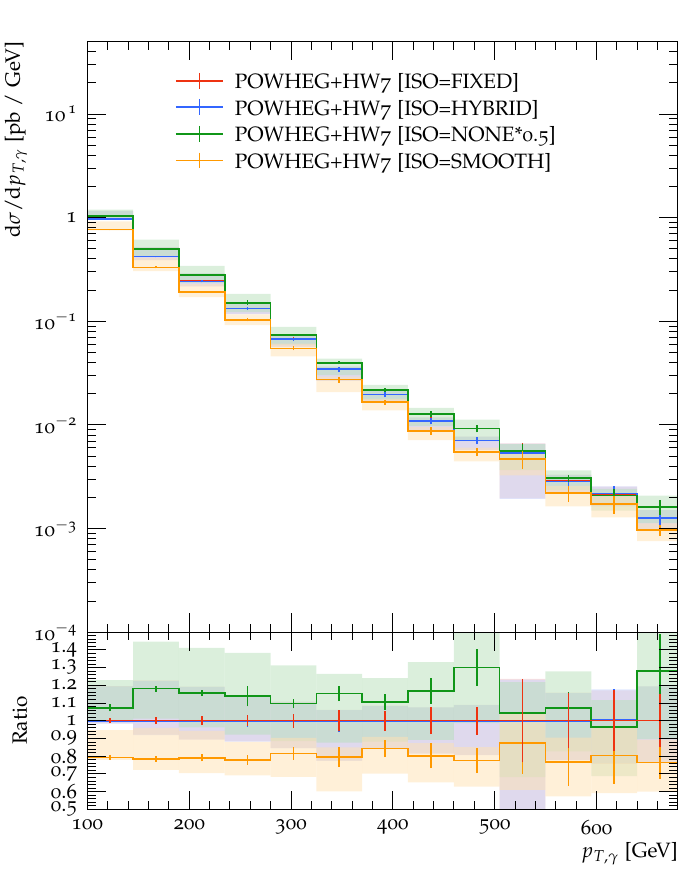}
    \caption{Transverse momentum distribution of the photon for different isolation schemes with no PS (left), \texttt{PYTHIA} (center) and \texttt{HERWIG} (right) parton showers.}
    \label{fig:pt}
\end{figure}

\begin{figure}
    \includegraphics[width=0.30\textwidth]{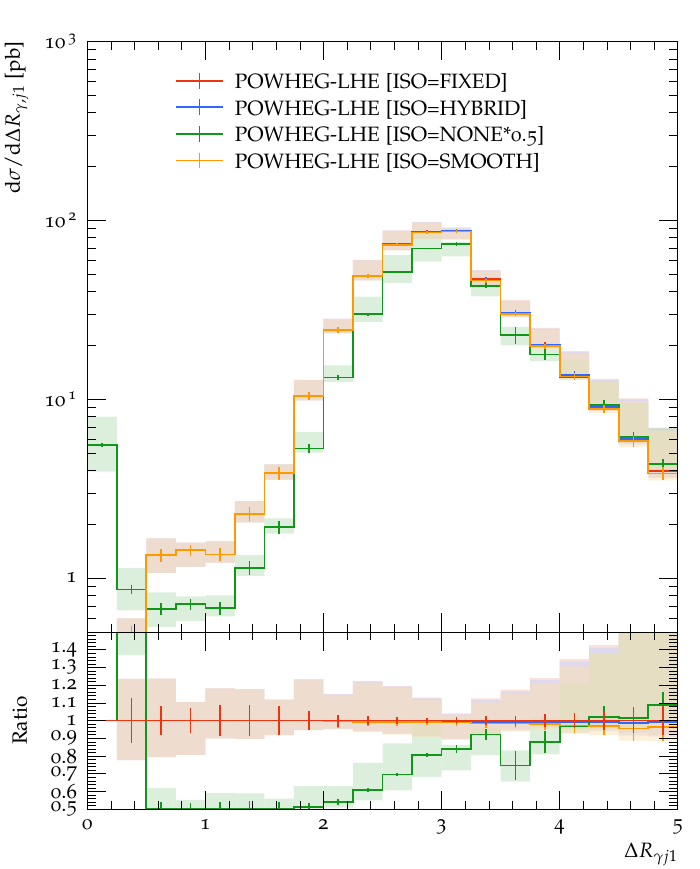}
    \includegraphics[width=0.30\textwidth]{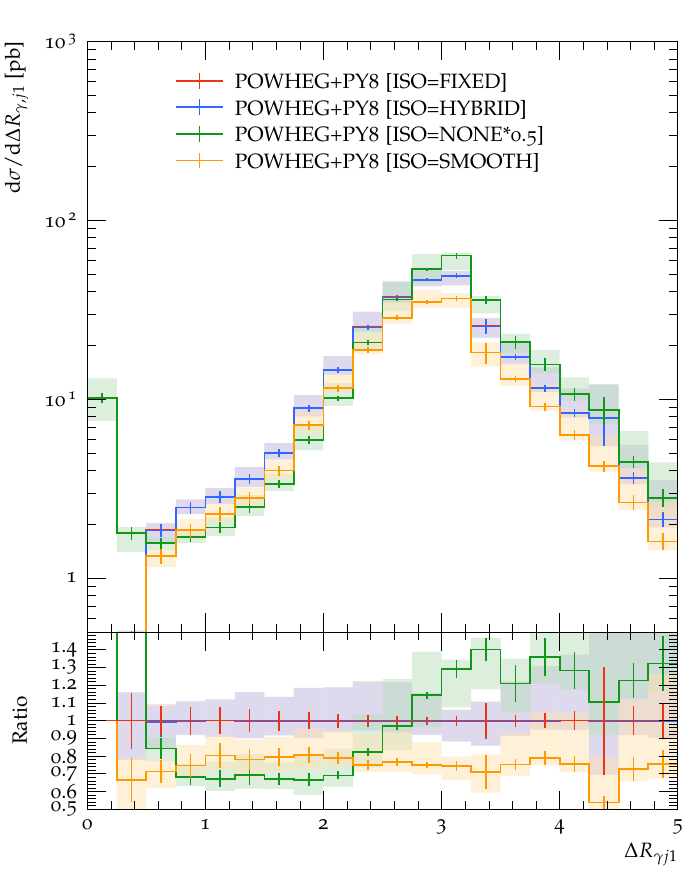}
    \includegraphics[width=0.30\textwidth]{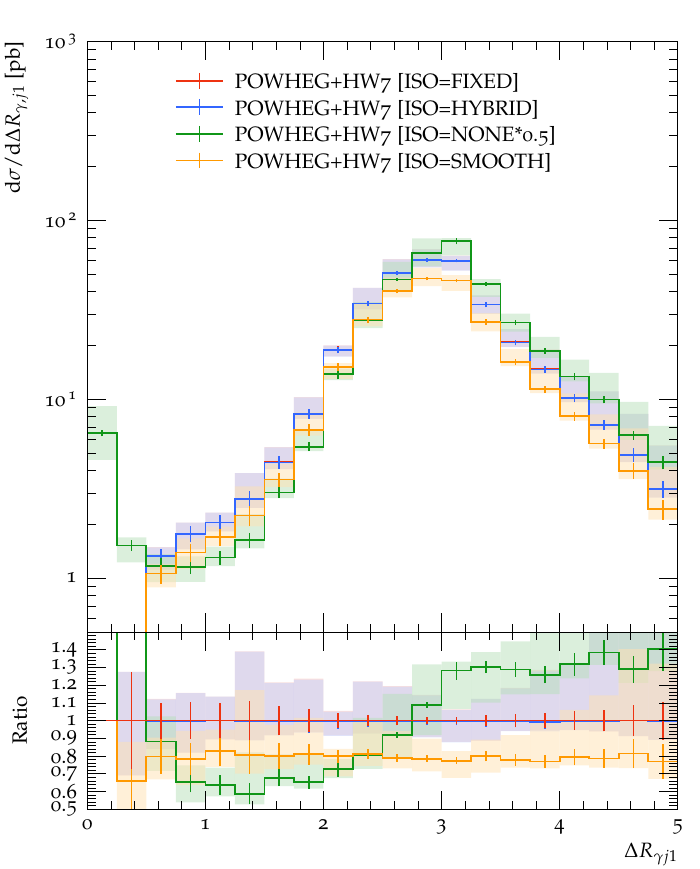}
    \caption{$\Delta R_{\gamma {\rm jet}_1}$ distribution of the photon for different isolation schemes using no PS (left), \texttt{PYTHIA} (center) and \texttt{HERWIG} (right).}
    \label{fig:dR}
\end{figure}

\begin{figure}
    \includegraphics[width=0.30\textwidth]{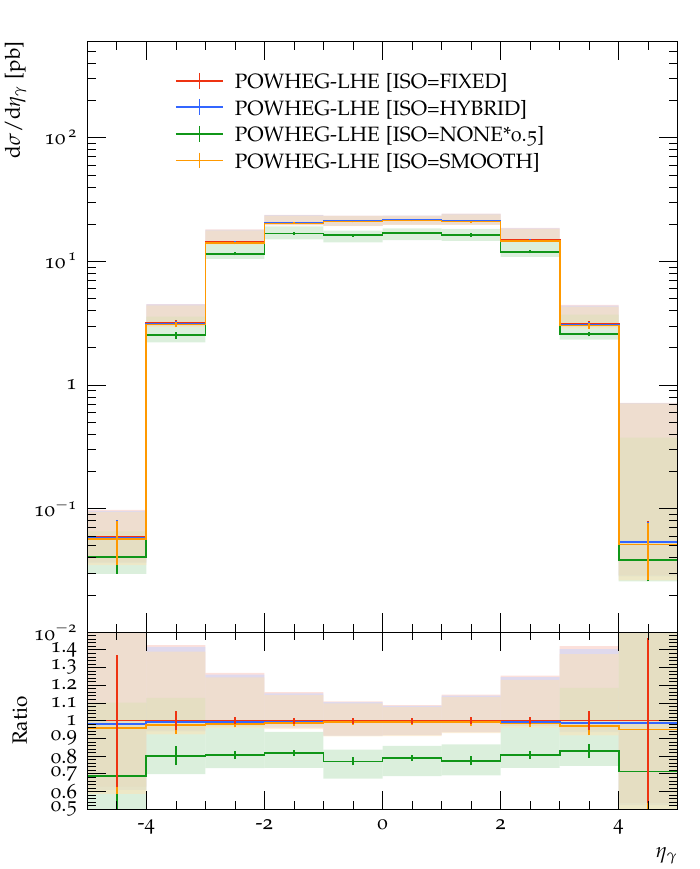}
    \includegraphics[width=0.30\textwidth]{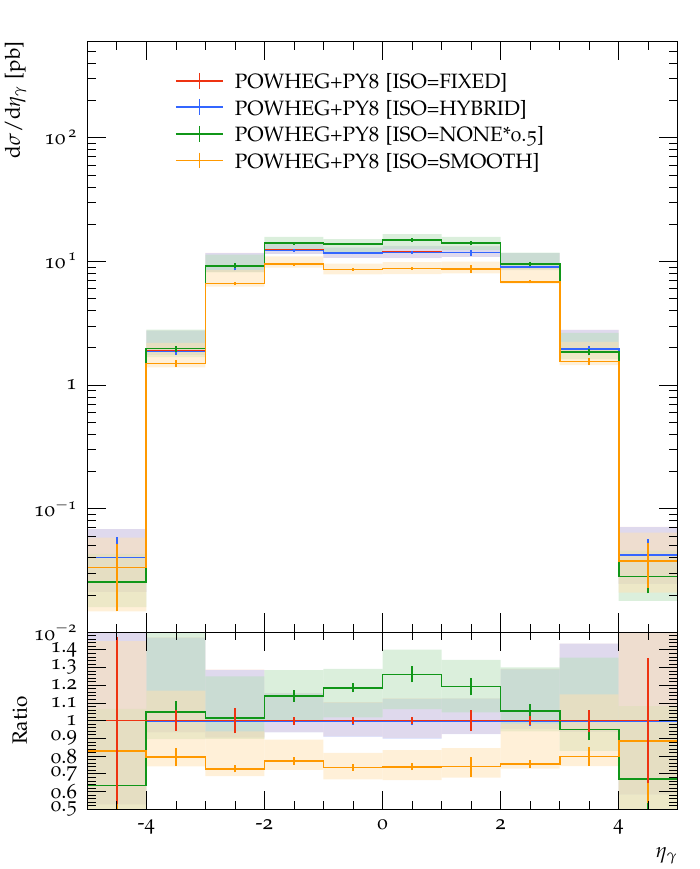}
    \includegraphics[width=0.30\textwidth]{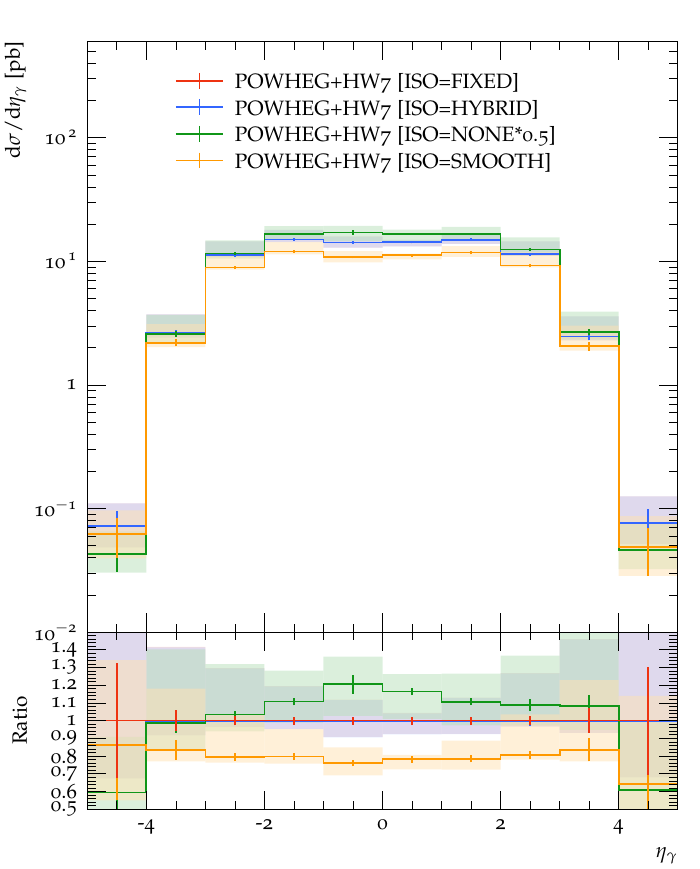}
    \caption{Rapidity distribution of the photon for different isolation schemes with no PS (left), \texttt{PYTHIA} (center) and \texttt{HERWIG} (right).}
    \label{fig:rapidity}
\end{figure}

In the study presented in Ref.~\cite{Amoroso:2020lgh}, three distinct photon isolation schemes are defined and compared, namely the fixed-cone, the smooth-cone, and the hybrid isolation. These can all be described by a single equation, which is then used to calculate the hadronic activity around the photon and to compare against the isolation energy
\begin{equation}
   \sum_{i}^{\Delta R_i < \epsilon R} p_T^i \left(\frac{\epsilon R}{r}\right)^{2n} \leq E_T^\text{max} = p_T^\text{iso}\,,
\end{equation}
where $(n=0,\epsilon=1)$ corresponds to the fixed-cone isolation, $(n=1,\epsilon=1)$ to the smooth-cone isolation, and the hybrid isolation requires both $(n=1,\epsilon=0.1)$ and $(n=0,\epsilon=1)$ to be satisfied. The fixed-cone isolation is the most common isolation scheme used in experimental analyses, as it is simple to implement and has a clear physical interpretation. Typically theoretical tools use the smooth-cone isolation, as it has the advantage that it does not get contributions from the parton-to-photon fragmentation functions \cite{Frixione:1998hn}. The hybrid isolation is a compromise between the two, as it is more restrictive than the fixed-cone isolation, but less restrictive than the smooth-cone isolation \cite{Siegert:2016bre,Chen:2019zmr}. Naturally, the hybrid isolation must result in cross sections that are lower than the fixed code isolation. This will only be barely visible in our setup (see below), however, and the hybrid isolation mostly results in the same prediction as the fixed isolation. Thus, it satisfies its purpose of addressing theoretical issues while remaining close to the experimentally motivated fixed-cone isolation. \Crefrange{fig:pt}{fig:rapidity} show our predictions for the different isolation schemes in \texttt{PY8} (left) and \texttt{HW7} (right). The parameters entering the analysis are the jet radius $R_j=0.4$ for the anti-$k_T$ algorithm~\cite{Cacciari:2008gp}, the isolation radius $R_I=0.4$ and the isolation energy $p_T^\text{iso}=2$ GeV. Contrary to the experimental analysis, in Ref.\ \cite{ATLAS:2019iaa}, there is neither a $p_T$ scaling of the isolation energy nor do we include a hadronic background subtraction, and we are only interested the hardest photon ($\gamma$) and hardest jet (jet$_1$). An implementation of our event selection in a \texttt{Rivet} analysis is available in the ancillary files accompanying this publication.\footnote{Alternatively, the source code is available at \url{https://gitlab.com/APN-Pucky/rivet-POWHEG_2023_DIRECTPHOTON}.}

\Cref{fig:pt} shows the distribution of the photon's transverse momentum. The red, yellow and blue curves correspond to the fixed-cone, smooth-cone, and hybrid isolation schemes, respectively. Additionally, in green the non-isolated contribution is shown and scaled by a factor of $1/2$, to facilitate the comparison of shapes. The change of shape as seen in the ratio panels between the non-isolated and fixed-cone isolation suggests that the isolation suppresses low $p_T$ photons more. For the smooth cone isolation this is even more pronounced, and the above parameters result in generally smaller cross sections than a fixed-cone isolation. There is, however, no significant difference in shape between the different isolation schemes, with mainly the smooth isolation being about 20\% lower than the fixed and hybrid isolation. Both parton showers reduce the cross section in comparison to their input, the unshowered events.  The LHE prediction alone shows no difference between the different isolation schemes. This can be explained by the fact that there are only single hard partons in the final state instead of a broader jet. The photon is then isolated if a parton is close to it and non-isolated otherwise. With a parton shower, however, the parton diffuses its energy into a cone-like jet with an increased probability to populate the photon's isolation cone continuously. The parton shower's reduction in the cross section is then partly due to the fact that less photons are isolated.

Next, in \Cref{fig:dR} we show how the different isolation schemes affect the separation between the photon and the hardest jet, jet$_1$. Comparing the non-isolated with the isolated curves one prominent difference in shape is the effective suppression above $\pi < \Delta R_{\gamma {\rm jet}_1} = \sqrt{(\Delta \eta)^2 + (\Delta \phi)^2}$. It is likely that even though the photon and the hardest jet are produced back-to-back, the photon is not isolated as the remaining jets can still become collinear with the photon. Then in the regime of $0.5< \Delta R_{\gamma {\rm jet}_1} < \pi$ the isolation reduces the cross section less. When the hardest jet is close to the isolation cone around the photon $ \Delta R_{\gamma j_1} < 0.5$ the isolation eliminates most of the cross section. Again, no difference in shape between the different isolation schemes is visible, with only the smooth isolation being about 20\%-30\% lower than the fixed and hybrid isolation. The production of soft emissions is visible in the increase in the low $\Delta R$ region between the no PS and parton shower figures.  A visible difference between the isolation schemes again exists only with the parton showers enabled.

In \Cref{fig:rapidity}, the rapidity distribution of the photon is shown. The interesting feature here is that \texttt{PY8} shows a larger suppression towards forward and backward rapidities than \texttt{HW7}, indicating a larger soft emission activity in these regions.
Without parton showers, the isolation scheme has no significant influence, but the influence is visible for smooth-cone isolation for both \texttt{PY8} and \texttt{HW7}. Coincidentally, the \texttt{HW7} results come here with slightly smaller numerical uncertainties and smaller fluctuations than the \texttt{PY8} results.

In general, we find that the shapes of the predictions do not depend significantly on the isolation criterion, but that the results for smooth-cone isolation lie below those for fixed-cone and hybrid isolation. The latter two agree in general quite well.

\FloatBarrier{}

\section{Conclusion}
\label{sec:conc}

In conclusion, we have presented a new calculation for prompt photon production in association with up to three jets at \gls*{NLO} in \gls*{QCD}, employing the POWHEG method for parton shower matching. By analyzing direct photon production with two jets and comparing these predictions with experimental ATLAS data, as well as \gls*{NNLO} calculations, we demonstrated that incorporating parton showers (both POWHEG+PYTHIA and POWHEG+HERWIG) significantly enhances the accuracy of the predictions. Notably, the NNLO predictions align well within the scale uncertainties with our POWHEG+PS simulations. The agreement with the experimental data provides support for the parton shower approach to describe the fragmentation process.
This new computation facilitates further studies of prompt photon production in association with multiple jets that will provide deep insights into the parton dynamics within nucleons and the properties of the \gls*{QGP}. Additionally, it lays the groundwork for implementing the \gls*{MiNLO} method in POWHEG for prompt photon production. However, it is important to recognize that the integration of \gls*{MiNLO} for prompt photons introduces additional complexities, particularly due to the necessity of carefully incorporating pure \gls*{QCD} contributions.

\acknowledgments
This work has been supported by the BMBF under contracts 05P21PMCAA and 05P24PMA and by the DFG through the Research Training Network 2149 “Strong and Weak Interactions - from Hadrons to Dark Matter”. Calculations for this publication were performed on the HPC cluster PALMA II of the University of Münster, subsidised by the DFG (INST 211/667-1).

\paragraph{Open Access.}
This article is distributed under the terms of the Creative Commons Attribution License (CC-BY-4.0), which permits any use, distribution and reproduction in any medium, provided the original authors(s) and source are credited.

\appendix

\section{MCFM-Rivet-YODA interface}
\label{app:mcfm}

During the course of our study, we developed an interface between MCFM-10.2.1~\cite{Campbell:2019dru}, Rivet-3~\cite{Bierlich:2019rhm} and YODA-1~\cite{Buckley:2023xqh}. The modified MCFM source code is available publically\footnote{\url{https://gitlab.com/APN-Pucky/mcfm}}. In \cref{lst:mcfm} we show the modifications to the MCFM nplotter to generate YODA histograms. Note, however, that the incoming and outgoing partons are misnamed as all gluons, since the MCFM nplotter does not provide the correct parton information (in fact it is already averaged). Thus the analysis should not rely on the parton information, but rather on the kinematics of the final state particles.

\begin{lstlisting}[caption=Additional code in MCFM nplotter in src/User/nplotter\_dirgam.f to generate YODA histograms., label=lst:mcfm]
    ids(1)=21
    ids(2)=21
    ids(3)=22
    ids(4)=21
    ids(5)=21
    npart = jets+3
    call to_rivet(p, ids,mxpart,npart,wt, analysis)
\end{lstlisting}

\section{POWHEG input}
\label{app:input}
\Cref{lst:input} shows the input file for the POWHEG event generation.
\begin{lstlisting}[caption=\texttt{POWHEG} input parameters, label=lst:input]
numevts 2000000   ! number of events to be generated
ih1   1           ! hadron 1 (1 for protons, -1 for antiprotons)
ih2   1           ! hadron 2 (1 for protons, -1 for antiprotons)
ebeam1 6500d0     ! energy of beam 1
ebeam2 6500d0     ! energy of beam 2
lhans1 27100      ! pdf set for hadron 1 (LHA numbering)
lhans2 27100      ! pdf set for hadron 2 (LHA numbering)

ncall1 100000   ! # calls for initializing the integration grid
itmx1    5      ! # iterations for initializing the integration grid
ncall2 100000   ! # calls for computing the integral and finding upper bounding envelope
itmx2    5      ! # iterations for computing the integral and finding upper bouding envelope
foldcsi   1     ! # folds on csi integration
foldy     1     ! # folds on  y  integration
foldphi   1     ! # folds on phi integration
nubound 100000  ! # calls to set up the upper bounding norms for radiation
icsimax  1      ! <= 100, number of csi subdivision when computing the upper bounds
iymax    1      ! <= 100, number of y subdivision when computing the upper bounds
xupbound 2d0    ! increase upper bound for radiation generation

bornktmin 90
bornsuppfact 1000 ! needed to get photons over a wide range of 100 GeV to over 1 TeV (ATLAS range)
suppnominlo 10

use-old-grid    1 ! if 1 use old grid
use-old-ubound  1 ! if 1 use norm of upper bounding function stored
novirtual   1   ! ignore virtuals, reenable in stage 4 (below)
nlotest     0
testplots 0
clobberlhe 1
smartsig 1
doublefsr  0  ! redefinition of emitter and emitted, recommended by powheg authors
iupperfsr  1  ! version of upperbounding, 2 ill-defined for massless process

iseed 10103        ! initialize random number sequence
manyseeds 1
maxseeds 2000
parallelstage 1
xgriditeration 1
compress_lhe 1     ! save disk space
compress_yoda 1    ! save disk space

alphaem_inv  127.0 ! 1/alphaem
renscfact  1.0d0   ! (default 1d0) ren scale factor: muren  = muref * renscfact
facscfact  1.0d0   ! (default 1d0) fac scale factor: mufact = muref * facscfact

!! shower param
NoPythia 0
QEDShower 1
QCDShower 1
hadronization 1
mpi 1
pythiaHepMC 1
HepMCEventOutput 0
HepMCEventInput 0
!! analysis param
NoRivet 0
rivetWeight 1 ! default weight, here 1 due to sudakov weight scaling
rivetAnalysis "ATLAS_2017_I1645627,ATLAS_2019_I1772071"
rivetAnalysis1 "MC_PHOTONINC,MC_PHOTONJETS,MC_PHOTONKTSPLITTINGS,MC_PHOTONS"
rivetAnalysis2 "POWHEG_2023_DIRECTPHOTON"
rivetAnalysis3 "POWHEG_2023_DIRECTPHOTON:ISO=SMOOTH"
rivetAnalysis4 "POWHEG_2023_DIRECTPHOTON:ISO=HYBRID"
rivetAnalysis5 "POWHEG_2023_DIRECTPHOTON:ISO=FIXED"
rivetAnalysis6 "POWHEG_2023_DIRECTPHOTON:ISO=NONE"

enhancedradfac 50 ! if > 0 then the splitting kernel used for photon radiation is multiplied by this factor (lhrwgt_id has to be set, too)
rwl_format_rwgt 1 ! HERWIG compatible LHE weights
rwl_file '-'
<initrwgt>
<weightgroup name='nominal' combine='None'>
<weight id='default'>default</weight>
<weight id='MUR1.0_MUF1.0_PDF27100'> emvirtual=1 novirtual=0 </weight>
</weightgroup>
<weightgroup name='scale_variations' combine='None'>
<weight id='MUR2.0_MUF2.0_PDF27100'> emvirtual=1 novirtual=0 renscfact=2d0 facscfact=2d0 </weight>
<weight id='MUR0.5_MUF0.5_PDF27100'> emvirtual=1 novirtual=0 renscfact=0.5d0 facscfact=0.5d0 </weight>
<weight id='MUR1.0_MUF2.0_PDF27100'> emvirtual=1 novirtual=0 renscfact=1d0 facscfact=2d0 </weight>
<weight id='MUR1.0_MUF0.5_PDF27100'> emvirtual=1 novirtual=0 renscfact=1d0 facscfact=0.5d0 </weight>
<weight id='MUR2.0_MUF1.0_PDF27100'> emvirtual=1 novirtual=0 renscfact=2d0 facscfact=1d0 </weight>
<weight id='MUR0.5_MUF1.0_PDF27100'> emvirtual=1 novirtual=0 renscfact=0.5d0 facscfact=1d0 </weight>
</weightgroup>
</initrwgt>
\end{lstlisting}

\section{Parton shower parameters}
\label{app:shower}

\Cref{lst:pythia} shows the parton shower parameters that differ from the defaults used in the \texttt{PYTHIA} simulation.
The value of \verb|pTmin| should be set to the same cutoff as in the POWHEG simulation.

\begin{lstlisting}[caption=\texttt{Pythia-8.306} parton shower parameters, label=lst:pythia]
Beams:frameType                               =                        5
Check:beams                                   =                      off
POWHEG:nFinal                                 =                        3
POWHEG:pTdef                                  =                        1
POWHEG:veto                                   =                        1
SpaceShower:pTmaxMatch                        =                        2
SpaceShower:pTmin                             =                  0.89443
TimeShower:pTmaxMatch                         =                        2
TimeShower:pTmin                              =                  0.89443
\end{lstlisting}

\Cref{lst:herwig} shows the parton shower parameters used in the \texttt{Herwig} simulation.

\begin{lstlisting}[caption=\texttt{Herwig-7.3.0} parton shower parameters,label=lst:herwig]
##################################################
#   Create the Les Houches file handler and reader
##################################################
cd /Herwig/EventHandlers
library LesHouches.so
# create the event handler
create ThePEG::LesHouchesEventHandler LesHouchesHandler

# set the various step handlers
set LesHouchesHandler:PartonExtractor /Herwig/Partons/PPExtractor

# set the weight option
set LesHouchesHandler:WeightOption VarNegWeight
set LesHouchesHandler:Weighted On

# set event hander as one to be used
set /Herwig/Generators/EventGenerator:EventHandler /Herwig/EventHandlers/LesHouchesHandler

# Set up an EMPTY CUTS object
# Normally you will have imposed any cuts you want
# when generating the event file and don't want any more
# in particular for POWHEG and MC@NLO you must not apply cuts on the
# the extra jet
create ThePEG::Cuts /Herwig/Cuts/NoCuts


# You can in principle also change the PDFs for the remnant extraction and
# multiple scattering. As the generator was tuned with the default values
# this is STRONGLY DISCOURAGED without retuning the MPI parameters
# create the reader and set cuts
create ThePEG::LesHouchesFileReader LesHouchesReader

set LesHouchesReader:AllowedToReOpen No
set LesHouchesReader:InitPDFs 0
set LesHouchesReader:Cuts /Herwig/Cuts/NoCuts

set LesHouchesReader:FileName pwgevents-0001.lhe.gz

# option to ensure momentum conservation is O.K. due rounding errors (recommended)
set LesHouchesReader:MomentumTreatment RescaleEnergy
# if using BSM models with QNUMBER info
#set LesHouchesReader:QNumbers Yes
#set LesHouchesReader:Decayer /Herwig/Decays/Mambo
# and add to handler
insert LesHouchesHandler:LesHouchesReaders 0 LesHouchesReader


####################################################################
# PDF settings #
####################################################################
# You may wish to use the same PDF as the events were generated with

set /Herwig/Partons/RemnantDecayer:AllowTop Yes

create ThePEG::LHAPDF /Herwig/Partons/PDFA ThePEGLHAPDF.so
set /Herwig/Partons/PDFA:PDFName MSHT20nlo_as118
set /Herwig/Partons/PDFA:RemnantHandler /Herwig/Partons/HadronRemnants
set LesHouchesReader:PDFA /Herwig/Partons/PDFA
create ThePEG::LHAPDF /Herwig/Partons/PDFB ThePEGLHAPDF.so
set /Herwig/Partons/PDFB:PDFName MSHT20nlo_as118
set /Herwig/Partons/PDFB:RemnantHandler /Herwig/Partons/HadronRemnants
set LesHouchesReader:PDFB /Herwig/Partons/PDFB
set /Herwig/Partons/PPExtractor:FirstPDF  /Herwig/Partons/PDFA
set /Herwig/Partons/PPExtractor:SecondPDF /Herwig/Partons/PDFB
cd /Herwig/EventHandlers
set LesHouchesHandler:HadronizationHandler /Herwig/Hadronization/ClusterHadHandler
set LesHouchesHandler:CascadeHandler /Herwig/Shower/ShowerHandler

##################################################
#  Shower parameters
##################################################
# normally, especially for POWHEG, you want
# the scale supplied in the event files (SCALUP)
# to be used as a pT veto scale in the parton shower
set /Herwig/Shower/ShowerHandler:MaxPtIsMuF Yes
set /Herwig/Shower/ShowerHandler:RestrictPhasespace Yes
# Shower parameters
# treatment of wide angle radiation
set /Herwig/Shower/PartnerFinder:PartnerMethod Random
set /Herwig/Shower/PartnerFinder:ScaleChoice Partner

set /Herwig/Shower/ShowerHandler:Interactions QEDQCD
read snippets/Rivet.in
insert /Herwig/Analysis/Rivet:Analyses 0 ATLAS_2019_I1772071
\end{lstlisting}

\bibliography{References}

\end{document}